\documentclass[a4paper, 11pt]{article}
\usepackage{mathrsfs}
\usepackage{enumitem}
\usepackage{hyperref}
\usepackage{color}
\usepackage{cite}
\usepackage{subfigure}
\usepackage{caption}
\usepackage[pdftex, a4paper]{geometry}
\usepackage{fancyhdr}
\usepackage{graphics}
\usepackage{amsfonts}
\usepackage{dsfont}
\usepackage{mathrsfs}
\usepackage{amssymb}
\usepackage{bbm}
\usepackage{epsfig}
\usepackage{amsmath}
\usepackage{appendix}
\usepackage[english]{babel}
\usepackage[all]{xy}

\newtheorem{theorem}{Theorem}
\newtheorem{definition}{Definition}

\newtheorem{lemma}{Lemma}
\newtheorem{remark}{Remark}
\newtheorem{proposition}{Proposition}

\begin{document}

\title{\bf Kalman Filtering over Gilbert-Elliott Channels: Stability Conditions  and the Critical Curve}
\date{}

\title{\bf Kalman Filtering over Gilbert-Elliott Channels: Stability Conditions  and the Critical Curve}
\date{}

\author{Junfeng Wu, Guodong Shi, Brian D. O. Anderson, and
Karl Henrik Johansson}

\maketitle

\begin{abstract}
This paper investigates  the stability of Kalman filtering
over Gilbert-Elliott channels where random packet drop follows a time-homogeneous two-state Markov chain whose state transition is determined by a pair of failure and recovery rates.
First of all, we establish  a relaxed  condition guaranteeing   peak-covariance stability
described by an inequality in terms of the spectral radius
of the system matrix and transition probabilities of
the Markov chain. We further show that
 that condition can be interpreted using a linear matrix inequality feasibility problem. Next, we prove that the peak-covariance stability implies mean-square stability, if the system matrix has no defective eigenvalues on the
unit circle. This connection between the two stability notions holds for any random packet drop process. {We prove that there exists a critical curve in the failure-recovery rate plane,  below which the Kalman filter is mean-square stable and no longer mean-square stable above, via a coupling method in stochastic processes.} Finally,  a lower bound for this critical failure rate is obtained  making use of the relationship we establish between the two stability criteria, based on an approximate relaxation of the system matrix.
\end{abstract}


\section{Introduction}

\subsection{Background and Related Works}
Wireless communications are being widely used nowadays in
sensor networks and networked control systems for a large spectrum of
applications, such as environmental monitoring, health care, smart building operation,
intelligent transportation and power grids.
New challenges accompany the considerable advantages wireless communications
offer in these applications, one of which is how channel fading and congestion,  influence the   performance of estimation and control. In the past decade, this fundamental  question has inspired various significant results  focusing on the interface of control and communication, and has become  a central theme in the study of networked sensor and control systems.

Early works on networked control systems assumed that sensors,
controllers, actuators and estimators communicate with each other over a finite-capacity
digital channel,~e.g., \cite{Delchamps, wong1, wong2, brockett1, nair1, tatikonda2,elia, Elia_N, ishii, fu-xie-tac05}, with the majority of contributions focused on one or both of finding
the minimum channel capacity or data rate needed for stabilizing the closed-loop
system, and constructing optimal encoder-decoder pairs to improve system
performance. At the same time, motivated by the fact that packets are the
fundamental information carrier in most modern data networks~\cite{joao07},
 many results on control or filtering with random packet dropouts appeared.

 State estimation, based on collecting measurements of the system output
from sensors deployed
in the field is embedded in many
networked control applications and is often
implemented recursively using a Kalman filter. Clearly, channel randomness leads to that the characterization of performance is not straightforward.
 A burst of interest in the problem of the stability of
Kalman filtering with intermittent measurements has arisen after
the pioneering work~\cite{Sinopoli2004}, where
{Sinopoli et al.} modeled the statistics of
intermittent observations by an
independent and identically distributed (i.i.d.)
Bernoulli random process and studied how packet losses affect
the state estimation.
It was proved that there exists a critical
arrival probability for packets, below which
the expected prediction error covariance matrix is no longer  uniformly bounded \cite{Sinopoli2004}.
Upper and lower bounds of this critical rate were provided for general systems, and it was shown that
the lower bound is tight for some special cases, such as when the observation matrix is invertible or
the system has a single unstable eigenvalue \cite{Sinopoli2004}. Further,
{Plarre and Bullo}~\cite{Plarre09tac} and
{Mo and Sinopoli}~\cite{yilin10bound} provided necessary and
sufficient conditions for the mean-square stability of a wider class of systems.
In~\cite{Plarre09tac}, it was shown that, when the system observation matrix
restricted to the observable subspace is invertible, the lower bound of the critical arrival probability
is tight. The result of~\cite{yilin10bound} revealed that, for
so-called non-degenerate
systems, the lower bound is also
sufficient. Results on the related problem of stabilization of closed-loop systems over packet lossy packet networks can be found
in~\cite{sinopoli2005optimal,imer2006optimalautomatica,
sinopoli2008optimal,foundation-ncs-packet-drops}.

To capture the temporal correlation of
realistic communication channels, the Gilbert-Elliott
model~\cite{gilbert1960capacity,elliott1963estimates}
that describes time-homogeneous Markovian packet losses has been introduced
to partially address this problem.~{Huang and Dey}~\cite{huang2006cdc,huang-dey-stability-kf} considered the
stability of Kalman filtering with Markovian packet losses.
To aid the analysis, they introduced a concept of peak covariance, defined by the expected prediction error covariance at the time instances when the channel just recovers from consecutive failed transmissions, as
an evaluation of estimation performance deterioration and
focused on its stability in the sense of its boundedness. Sufficient conditions for the peak-covariance stability were
proposed for general vector systems with  a necessary and sufficient
condition for scalar systems, and
the relationship between the mean-square stability and the peak-covariance stability
was discussed~\cite{huang-dey-stability-kf}. Improvements  to these results appeared in
~\cite{xie2007peak,xie2008stability}.
Parallel to this, in~\cite{you2011mean},
by investigating the estimation error covariance matrices at
each packet reception time,
necessary and sufficient conditions for the mean-square stability
were derived for second-order systems and certain classes of higher-order
systems. It is intuitive
that the time instants at which the channel just recovers
are instants when the covariance might be at maximum, given that
when all packets are lost, the covariance is always growing, but this
maximum property was never actually established in the references above.
Essentially, the probabilistic characteristics of the prediction error covariance are fully captured by
its probability distribution function. Motivated by this,
{Shi et al.}~\cite{shi-tac10} studied Kalman filtering with random packet losses
from a probabilistic perspective where the
performance metric was defined using the error covariance matrix distribution function, instead of the mean. {Mo and Sinopoli}~\cite{yilin12criticalvalue}
studied the decay rate of the estimation error covariance matrix,
and derived the critical arrival probability for non-degenerate systems based on the decay rate.
Weak convergence of Kalman filtering with packet losses, i.e.,
that error covaraince matrix  converges to a limit distribution, were investigated in  \cite{kar2012kalman,censi2011kalman,xie2012stochastic}  for
i.i.d., semi-Markov, and Markov drop models,  respectively.

\subsection{Contributions and Paper Organization}
In this paper, we focus on the peak-covariance and mean-square stabilities  of
Kalman filtering with Markovian packet losses.
The motivation is from our observation that the existing literature~\cite{huang2006cdc,huang-dey-stability-kf,xie2007peak,xie2008stability} is incomplete, as restrictive assumptions are made on the plant dynamics and the communication channel.
We show by numerical examples that
conditions for peak-covariance stability in the literature
only apply to reliable channels with low failure
rate. Moreover, existing results
rely on calculating
an infinite sum of matrix norms in the checking of stability conditions. Although it was proved  that with i.i.d. packet losses the
peak-covariance stability is equivalent to the mean-square stability
for scalar systems and systems that are one-step observable~\cite{huang-dey-stability-kf,xie2008stability},
for vector systems with more general  packet drop processes,
this relationship is yet unclear.
In this paper, we first derive relaxed and explicit  peak-covariance stability conditions. Then we establish a result indicating that peak-covariance stability implies mean-square stability under quite general settings. We eventually make use of these results to obtain mean-square stability criteria.
The contributions of this paper are summarized as follows.
\begin{itemize}
\item A relaxed  condition guaranteeing   peak-covariance stability is obtained
described by an inequality in terms of the spectral radius
of the system matrix and transition probabilities of
the Markov chain, rather than an infinite sum of matrix norms
as in~\cite{huang2006cdc,huang-dey-stability-kf,xie2007peak,xie2008stability}. We show that
 that condition can be recast as a linear matrix inequality (LMI) feasibility problem. These conditions are theoretically and numerically shown to be less conservative
than those in the literature.

\item We prove that peak-covariance stability implies mean-square stability if the system matrix has no defective eigenvalues on the
unit circle. Remarkably enough this implication holds for any random packet drop process that allows peak-covariance stability to be defined. This result bridges two stability criteria in the literature, and offers a tool for studying mean-square stability of the Kalman filter
 through its peak-covariance stability. Note that mean-square stability
 was previously studied using quite different methods such as
analyzing the boundness of the expectation of a kind of randomized
observability Gramians over a stationary random packet loss process
to establish the equivalence between stability in stopping times and stability
in sampling times~\cite{you2011mean}, and characterizing the decay rate of
the prediction covariance's tail distribution for so-called non-degenerate
systems~\cite{yilin12criticalvalue}.

\item  We further prove that
for a fixed recovery rate in the transition probability matrix,
there exists a critical failure rate  such that
if and only if the failure rate is below the critical value
the expected prediction error covariance matrices
are uniformly bounded. Let the failure rate $p$ and
recovery rate $q$ define a Gilbert-Elliott channel.
It is shown that there exists a critical region
in the $p-q$ plane such that
if and only if the pair $(p,q)$
falls into that region the expected prediction error covariance matrices
are uniformly bounded.
Finally, we present a lower bound for the critical failure rate, making use of  the relationship between the two stability criteria we established. This lower bound holds without relying on the restriction that the system matrix has no defective eigenvalues on the unit circle. In other words, we obtain a mean-square stability condition for \emph{general} linear time-invariant (LTI) systems under Markovian packet drops.
\end{itemize}
We believe these results add to the fundamental understanding of Kalman filtering under random packet drops.

The remainder of the paper is organized as follows. Section~\ref{section:problem-setup} presents the problem setup.
Section~\ref{section:peak-covariance-Stability}
focuses on the peak-covariance stability.
Section~\ref{section:mss-KF} studies the relationship between the peak-covariance
and mean-square stabilities, the critical $p-q$ curve, and presents
 a sufficient condition
for mean-square stability of general LTI systems.
Two numerical examples
in Section~\ref{section:simulation} demonstrate the effectiveness of our
approach compared with the literature.
Finally we provide some concluding remarks in Section~\ref{section:conclusions}.

\textit{Notations}: $\mathbb{N}$ is
the set of positive integers.
$\mathbb{S}_{+}^{n}$ is the set of $n$ by $n$ positive
semi-definite matrices over the complex field. For a matrix $X$,
$\sigma(X)$ denotes the spectrum of $X$
and $\lambda_X$ denotes the eigenvalue of $X$ that has the largest magnitude.
$X^*$, $X'$ and $\overline X$ are the Hermitian conjugate, transpose and complex conjugate
of $X$.
Moreover, $\|\cdot\|$ means the 2-norm of a vector or the induced 2-norm of a matrix.
$\otimes$ is the Kronecker product of two matrices.
The indicator function of a subset $\mathcal{A}\subset\Omega$ is a function
${1}_\mathcal{A}:\Omega \rightarrow \{ 0,1 \}$, where
${1}_\mathcal{A}(\omega)=1$ if $\omega\in \mathcal{A}$, otherwise
${1}_\mathcal{A}(\omega)=0$. For random variables, $\sigma(\cdot)$ is
the $\sigma$-algebra generated by
the variables.

\section{Kalman Filtering with Markovian Packet Losses}\label{section:problem-setup}

Consider an LTI system:
\begin{eqnarray}
x_{k+1} & = & Ax_k + w_k, \label{eqn:process-dynamics} \\
y_k & = & Cx_k + v_k, \label{eqn:measurement-equation}
\end{eqnarray} where $A\in \mathbb{R}^{n\times n}$ is the system matrix and $C\in \mathbb{R}^{m\times n}$ is the observation matrix, $x_k \in \mathbb{R}^{n}$ is the process state vector and $y_k \in \mathbb{R}^{m}$ is the observation vector, $w_{k} \in \mathbb{R}^{n} $ and
$v_k \in \mathbb{R}^{m}$ are zero-mean Gaussian random vectors
with auto-covariance  $\mathbb{E}[w_{k}w_{j}{'}] = \delta_{kj}Q~(Q\geq 0)$,
$\mathbb{E}[v_{k}v_{j}{'}] = \delta_{kj}R~(R > 0)$,
$\mathbb{E}[w_{k}v_{j}{'}] = 0 \; \forall j,k$. Here $\delta_{kj}$ is the Kronecker
delta function with $\delta_{kj} = 1$ if $k=j$ and $\delta_{kj}=0$ otherwise.
The initial state $x_0$ is a zero-mean Gaussian random vector that is uncorrelated
with $w_k$ and $v_k$ and has covariance $\Sigma_0\geq 0$.
We assume that $(C,A)$ is detectable and $(A,Q^{{1}/{2}})$ is stabilizable.
By applying a similarity transformation, the unstable and stable modes of the considered  LTI system can be decoupled. An open-loop prediction of the stable mode always has a bounded estimation error covariance, therefore, this mode does not play any key role in the stability issues  considered in this paper.
Without loss of generality, we assume that
\begin{enumerate}[label={(A\arabic*)}]
\item\label{asmpt:assumpt-A}
\textit{All of the eigenvalues of $A$ have magnitudes not less than one.}
\end{enumerate}
Certainly $A$ is nonsingular, $(C,A)$ is observable and $(A,Q^{{1}/{2}})$ is controllable.

We consider an estimation scheme
where the raw measurements of the sensor $\{y_k\}_{k\in\mathbb{N}}$
are transmitted to the estimator via an erasure communication channel over which packets may be dropped randomly.
Denote by $\gamma_k\in\{0,1\}$ the arrival of $y_k$ at time $k$:
 $y_k$ arrives error-free at the estimator if $\gamma_k=1$; otherwise
$\gamma_k=0$.
Whether $\gamma_k$ takes value $0$ or $1$ is assumed to be known by the
receiver at time $k$.
Define $\mathcal{F}_k$ as the filtration generated by all the measurements received by
the estimator up to time $k$, i.e.,
$\mathcal{F}_{k}\triangleq \sigma\hspace{-0.5mm}\left
(\gamma_ty_t,\gamma_t;1\leq t\leq k\right)$ and
$\mathcal{F}=\sigma\left(\cup_{k=1}^\infty \mathcal{F}_k\right)$.
We will use a triple $(\Omega,\mathcal{F},\mathbb{P})$ to
denote the probability space capturing all the randomness in the model.

To describe the temporal correlation of
realistic communication channels, we assume the Gilbert-Elliott channel \cite{gilbert1960capacity,elliott1963estimates}, where
the packet loss process is a time-homogeneous two-state Markov
chain. To be precise,  $\{\gamma_k\}_{k\in \mathbb{N}}$ is the state of the Markov chain with initial condition,
without loss of generality, $\gamma_1=1$.
The transition probability matrix for the Gilbert-Elliott channel is given by
\begin{equation}\label{eqn:markov-trnsition-prob}
\mathbf{P}=\left[\begin{array}{ccc}
1-q & q \\
p & 1-p \\
\end{array}\right],
\end{equation}
where $p\triangleq\mathbb{P}(\gamma_{k+1}=0|\gamma_k=1)$ is called the failure rate, and $q\triangleq\mathbb{P}(\gamma_{k+1}=1|\gamma_k=0)$ is called the recovery rate. Assume that
\begin{enumerate}[label={(A2)}]
\item\label{asmpt:assumpt-A2}
\textit{The failure and recovery rates satisfy $p,q\in(0,1)$.}
\end{enumerate}
Then this  Markov chain is ergodic and has the unique stationary distribution
$$
\lim_{k\rightarrow \infty}\mathbb{P}(\gamma_k=1)=\frac{q}{p+q},~~\;\;\;
\lim_{k\rightarrow \infty}\mathbb{P}(\gamma_k=0)=\frac{p}{p+q}.
$$

The estimator computes $\hat{x}_{k|k}$, the minimum mean-squared error estimate,
and $\hat x_{k+1|k}$, the one-step prediction, according to
$\hat{x}_{k|k} =\mathbb{E}[x_k|\mathcal{F}_k]$ and
$\hat{x}_{k+1|k} =\mathbb{E}[x_{k+1}|\mathcal{F}_k]$.
 Let $P_{k|k}$ and $P_{k+1|k}$ be the corresponding estimation and prediction error covariance matrices, i.e.,
$P_{k|k}=\mathbb{E}[(x_k-\hat x_{k|k}) (\cdot)'|\mathcal{F}_k]$
and $P_{k+1|k}=\mathbb{E}[(x_{k+1}-\hat x_{k+1|k}) (\cdot)'|\mathcal{F}_k]$.
They can be computed recursively via a modified Kalman filter~\cite{Sinopoli2004}.
The recursions for $\hat{x}_{k|k}$ and $\hat{x}_{k+1|k}$ are omitted here. To study the Kalman filtering system's stability, we focus
on the the prediction error covariance matrix $P_{k+1|k}$, which is recursively
computed as
\begin{equation*}
  P_{k+1|k} =  AP_{k|k-1}A' + Q-\gamma_k AP_{k|k-1}C'
  (CP_{k|k-1}C'+R)^{-1}CP_{k|k-1}A'.
\end{equation*}
It can be seen that $P_{k+1|k}$ inherits the randomness of $\{\gamma_t\}_{1\leq t
\leq k}$. In what follows, we focus on
characterizing the impact of $\{\gamma_k\}_{k\in {\mathbb{N}}}$ on $P_{k+1|k}$.
To simplify notations in the sequel, let
$P_{k+1}\triangleq P_{k+1|k}$, and
define the functions $h$, $g$, $h^k$ and $g^k$: $\mathbb{S}^n_+ \to \mathbb{S}^n_+$ as
follows:
\begin{eqnarray}
h(X)&\triangleq& AXA'+Q,\label{eqn:h-func}\\
g(X)&\triangleq &AXA'+Q-AXC'
{(CXC'+R)^{-1}CXA'},\label{eqn:g-func}
\end{eqnarray}\vspace{2mm}
$h^k(X)\triangleq \underbrace{h\circ h \circ \cdots \circ h}_{k \hbox{~times}}(X)$ and
$g^k(X)\triangleq \underbrace{g\circ g \circ \cdots \circ g}_{k \hbox{~times}}(X)$,
where $\circ$ denotes the function composition.

\section{Peak-covariance Stability}\label{section:peak-covariance-Stability}
In this section, we study the peak-covariance stability~\cite{huang-dey-stability-kf} of the Kalman filter. To this end, we define
\begin{eqnarray}\label{def:stopping-times}
\tau_1&\triangleq& \min\{k:k\in\mathbb{N}, \gamma_k=0\},\notag\\
\beta_1&\triangleq& \min\{k: k>\tau_1, \gamma_k=1\},\notag\\
&\vdots&\notag\\
\tau_j&\triangleq& \min\{k: k>\beta_{j-1}, \gamma_k=0\},\notag\\
\beta_j&\triangleq& \min\{k: k>\tau_{j}, \gamma_k=1\}.\label{def:beta-j}
\end{eqnarray}
It is straightforward to verify that $\{\tau_j\}_{j\in\mathbb{N}}$ and $\{\beta_j\}_{j\in\mathbb{N}}$ are two sequences of stopping times because
both $\{\tau_j\leq k\}$ and $\{\beta_j\leq k\}$ are $\mathcal{F}_k-$measurable;
see~\cite{durrett2010probability} for details. Due to the strong Markov property and the ergodic property
of the Markov chain defined by~\eqref{eqn:markov-trnsition-prob} (e.g., \cite{huang-dey-stability-kf}), the sequences $\{\tau_j\}_{j\in\mathbb{N}}$ and $\{\beta_j\}_{j\in\mathbb{N}}$
have finite values $\mathbb{P}$-almost surely.
Then we can define the sojourn times at the state $1$ and state $0$ respectively
by $\tau_j^*$ and $\beta_j^*~\forall j\in\mathbb{N}$ as
\begin{eqnarray*}
\tau_j^*&\triangleq& \tau_j-\beta_{j-1},\\
\beta_j^*&\triangleq& \beta_j-\tau_j,
\end{eqnarray*}
where we define $\beta_0=1$. The following result given by~\cite{huang-dey-stability-kf} demonstrates that
$\{\tau_k^*\}_{j\in\mathbb{N}}$ and $\{\beta_k^*\}_{j\in\mathbb{N}}$ are i.i.d. and mutually independent.
\begin{lemma}[Lemma 2 in~\cite{huang-dey-stability-kf}]\label{lemma:stopping-time-lemma}
Under~(A2), the following statements on $\{\tau_j^*\}_{j\in\mathbb{N}}$ and $\{\beta_j^*\}_{j\in\mathbb{N}}$ hold:
\begin{enumerate}
\item[(i).] $\{\tau_j^*\}_{j\in\mathbb{N}}$ is an i.i.d. random sequence, and $\tau_j^*-1$ is
geometrically distributed, i.e., \\ $\mathbb{P}\left(\tau_j^*-1=k\right)=(1-p)^kp$, $k\in\mathbb{N}$;

\item[(ii).] $\{\beta_j^*\}_{j\in\mathbb{N}}$ is an i.i.d. random sequence, and $\beta_j^*-1$ is
geometrically distributed, i.e., \\ $\mathbb{P}\left(\beta_j^*-1=k\right)=(1-q)^kq$, $k\in\mathbb{N}$;

\item[(iii).] $\tau_1^*,\beta_1^*,\ldots,\tau_j^*,\beta_j^*,\ldots$ define a sequence of independent random variables.
\end{enumerate}
\end{lemma}
Let us denote the prediction error covariance matrix at the stopping time $\beta_j$ by $P_{\beta_j}$ and call it the peak covariance\footnote{
The definition of peak covariance was first introduced in~\cite{huang-dey-stability-kf}, where the term ``peak'' was attributed to the fact that  for an unstable scalar system $P_k$ monotonically increases to reach a local maximum at time $\beta_j$. This maximum property does not
necessarily hold for the multi-dimensional case.} at $\beta_j$.
To study the stability of Kalman filtering with Markovian packet losses, we introduce the
concept of peak-covariance stability~\cite{huang-dey-stability-kf}, as follows:
\begin{definition}\label{def:peak-cov-stability}
The Kalman filtering system with packet losses is said to be peak-covariance stable if $\sup_{j\in\mathbb{N}} \mathbb{E} \| P_{\beta_j}\|<\infty$.
\end{definition}
\subsection{Stability Conditions}
To analyze the peak-covariance stability, we introduce
the observability index of the pair $(C,A)$.
\begin{definition}\label{def:observability-index}
The observability index $\mathsf{I_o}$ is defined as the
smallest integer such that
\\$[C',A'C',\ldots,(A^{{I}_o-1})'C']'$ has rank $n$.
If $\mathsf{I_o}=1$, the system $(C,A)$ is called one-step observable.
\end{definition}

We have the following result.

\begin{theorem}\label{thm:main-thm}
Suppose the following two conditions hold:
\begin{enumerate}
\item[(i).] $|\lambda_A|^2(1-q)<1$;
\item[(ii).]  $\exists\,K\triangleq[K^{(1)},\ldots,K^{(\mathsf{I_o}-1)}]$, where $K^{(i)}$'s are matrices with compatible dimensions, such that $|\lambda_{H(K)}|<1$, where
\begin{equation}\label{eqn:main-sufficient-cond}
H(K)=qp\Big[(A\otimes A)^{-1}-(1-q)I\Big]^{-1}\sum_{i=1}^{\mathsf{I_o}-1}
\overline{(A^i+K^{(i)}C^{(i)})}\otimes (A^i+K^{(i)}C^{(i)})(1-p)^{i-1}.
\end{equation}
\end{enumerate}
Then $\sup_{j\geq 1}\mathbb{E}\|P_{\beta_j}\|<\infty$, i.e., the Kalman filtering system
is peak-covariance stable.
\end{theorem}
\begin{remark}
In~\cite{you2011mean}, the authors defined stability in stopping times as the stability of $P_k$ at packet
reception times. Note that $\{\beta_j\}_{j\in\mathbb{N}}$, at which the peak covariance is defined, can also be treated as the stopping times defined
on packet reception times. Clearly, in scalar systems, the covariance is at maximum
when the channel just recovers from failed transmissions; therefore
peak covariance sequence gives an upper envelop of covariance matrices at packet reception times. For higher-order systems, the relation between them is still unclear.
\end{remark}
Theorem~\ref{thm:main-thm} is proved via investigating the vectorization of $P_{\beta_k}$, and the detailed proof is given in Appendix B. As the second condition in Theorem~\ref{thm:main-thm} is difficult to directly verify, in the following proposition we present  another condition for peak-covariance stability, which is, despite being conservative, easy to check. The new condition is obtained by making all $K^{(i)}$'s in Theorem~\ref{thm:main-thm} take the value zero.
\begin{proposition}\label{corollary:main-result}
 If the following condition is satisfied:
\begin{equation}\label{eqn:sufficient-cond-2}
pq|\lambda_A|^2\sum_{i=1}^{\mathsf{I_o}-1}|\lambda_A|^{2i}(1-p)^{i-1}<1-|\lambda_A|^2(1-q),
\end{equation}
then the Kalman filtering system is peak-covariance stable.
\end{proposition}
{\it Proof.}
The proof requires the following lemma.
\begin{lemma}[Theorem 1.1.6 in~\cite{horn2012matrix}]\label{lemma:eigenvalue-polynomial}
Let $\mathfrak{p}(\cdot)$ be a given polynomial. If $\lambda$ is an eigenvalue of
a matrix $A$, then $\mathfrak{p}(\lambda)$ is an eigenvalue of the matrix
$\mathfrak{p}(A)$.
\end{lemma}
Define a sequence of polynomials of the matrix $A\otimes A$ as $\{\mathfrak{p}_n(A\otimes A)\}_{n\in\mathbb{N}}$, where
$$
\mathfrak{p}_n(A\otimes A)= \sum_{i=1}^n(A\otimes A)^i(1-q)^{i-1}q\sum_{j=1}^{\mathsf{I_o}-1}(A\otimes A)^j(1-p)^{j-1}p.
$$
In light of Lemma~\ref{lemma:eigenvalue-polynomial},  the spectrum of $\mathfrak{p}_n(A\otimes A)$ is given by
$\sigma\left(\mathfrak{p}_n(A\otimes A)\right)=\{\mathfrak{p}_n(\lambda_i\lambda_j):~\lambda_i,\lambda_j\in \sigma(A)\}.$ Since
$A$ is a real matrix, its complex eigenvalues, if any, always occur in conjugate pairs.
Therefore, $|\lambda_A|^2$ must be an eigenvalue of
$A\otimes A$, and
the spectral radius of $\mathfrak{p}_n(A\otimes A)$ can be computed as
\begin{eqnarray*}
|\lambda_{\mathfrak{p}_n(A\otimes A)}|=\sum_{i=1}^n
|\lambda_A|^{2i}(1-q)^{i-1}q\sum_{j=1}^{\mathsf{I_o}-1}|\lambda_A|^{2j}(1-p)^{j-1}p.
\end{eqnarray*}
It is evident that the sequence $\{|\lambda_{\mathfrak{p}_n(A\otimes A)}|\}_{n\in\mathbb{N}}$ is
monotonically increasing. When $|\lambda_A|^2(1-q)<1$, we have
\begin{equation}\label{eqn:H0}
\lim_{n\rightarrow \infty}\mathfrak{p}_n(A\otimes A)=H(0)
\end{equation}
and
\begin{equation}\label{eqn:spectrum-radius-H0}
\lim_{n\rightarrow \infty}|\lambda_{\mathfrak{p}_n(A\otimes A)}|=
\frac{q|\lambda_A|^2}{1-|\lambda_A|^2(1-q)}\sum_{j=1}^{\mathsf{I_o}-1}|\lambda_A|^{2j}(1-p)^{j-1}p.
\end{equation}
As $|\lambda_X|$ is continuous with respect to $X$,~\eqref{eqn:H0} and~\eqref{eqn:spectrum-radius-H0} altogether lead to
$$|\lambda_{H(0)}|=\frac{q|\lambda_A|^2}{1-|\lambda_A|^2(1-q)}\sum_{j=1}^{\mathsf{I_o}-1}|\lambda_A|^{2j}(1-p)^{j-1}p.$$
Letting $K^{(i)}=0~\forall 1\leq i\leq \mathsf{I_o}-1$, the condition
provided in Theorem~\ref{thm:main-thm} becomes:
$(i).$~$|\lambda_A|^2(1-q)<1$, $(ii).$~$|\lambda_{H(0)}|<1$. Since the left side
of~\eqref{eqn:sufficient-cond-2} is nonnegative, it imposes the
positivity of $1-|\lambda_A|^2(1-q)$, whereby the conclusion follows.
\hfill$\square$
\begin{remark}
The left side of~\eqref{eqn:sufficient-cond-2} is strictly positive when $\mathsf{I_o}\geq 2$, while it
vanishes when $\mathsf{I_o}=1$. In the latter case, plus the necessity as shown in~\cite{xie2007peak}, $|\lambda_A|^2(1-q)<1$ thereby becomes a necessary and sufficient condition
for peak-covariance stability. This observation is consistent with
the conclusion of Corollary 2 in~\cite{xie2008stability}.
\end{remark}

Theorem~\ref{thm:main-thm} and Proposition~\ref{corollary:main-result} establish a direct connection between $\lambda_A$ (or $\lambda_{H(K)}$), $p$, $q$,
the most essential aspects of the system dynamic and channel characteristics on the one hand,
and peak-covariance stability on the other hand.
These results cover the ones in~\cite{huang2006cdc,huang-dey-stability-kf,xie2007peak,xie2008stability}, as is evident using the subadditivity property of matrix norm, and the fact that the spectral radius is the infimum of all possible matrix norms.
To see this, one should notice that
\begin{eqnarray*}
|\lambda_{H(k)}|&\leq& qp\Big\|\Big[(A\otimes A)^{-1}-(1-q)I\Big]^{-1}\Big\|\sum_{i=1}^{\mathsf{I_o}-1}
(1-p)^{i-1}\left\|\overline{(A^i+K^{(i)}C^{(i)})}\otimes (A^i+K^{(i)}C^{(i)})\right\|\\
&\leq&qp \sum_{i=1}^\infty (1-q)^{i-1}\|A^i\otimes A^i\|
\sum_{i=1}^{\mathsf{I_o}-1}
(1-p)^{i-1}\left\|\overline{(A^i+K^{(i)}C^{(i)})}\otimes (A^i+K^{(i)}C^{(i)})\right\|\\
&=&q\sum_{i=1}^\infty (1-q)^{i-1}\|A^i\|^2
p\sum_{i=1}^{\mathsf{I_o}-1}
(1-p)^{i-1}\|A^i+K^{(i)}C^{(i)}\|^2,
\end{eqnarray*}
in which the first inequality follows from $|\lambda_{H(K)}|\leq \|H(K)\|$
and the submultiplicative property of matrix norms, and
the last equality holds because, for a matrix $X$, $\|\overline{X}^i\otimes X^i\|=\sqrt{\lambda_{(\overline {X}^i(X')^i)\otimes (X^i(X^*)^i)}}=\lambda_{X^i(X^*)^i}=\|X^i\|^2$.
Comparison with the related results in the literature is also demonstrated by
Example I in Section~\ref{section:simulation}.

\subsection{LMI Interpretation}
In Theorem~\ref{thm:main-thm},
a quite heavy computational overhead may be incurred in searching for a satisfactory $K$.
Although computationally-friendly, Proposition~\ref{corollary:main-result} only provides
a comparably rough criterion. In this part, we continue to polish the result of Theorem~\ref{thm:main-thm} with a way to retaining its power but with less computational burden.
Based on what we have established in Theorem~\ref{thm:main-thm}, we present a criterion
which reduces to solving an LMI feasibility problem.
To do so, we first introduce a linear operator and then
present the equivalence between several statements related to
this linear operator (including an LMI feasibility statement) and
$|\lambda_{H(K)}|<1$, any of which results in the peak-covariance stability.

Consider the operator
$\mathcal{L}_K:\mathbb{S}_+^n\rightarrow\mathbb{S}_+^n$ defined as
\begin{equation}\label{eqn:L_K_X}
\mathcal{L}_K(X)=p\sum_{i=1}^{\mathsf{I_o}-1}(1-p)^{i-1}(A^i+K^{(i)}C^{(i)})^*
\Phi_X(A^i+K^{(i)}C^{(i)}),
\end{equation}
where $\Phi_X$ is the positive definite solution of the Lyapunov equation
$(1-q)A'\Phi_X A+qA'XA=\Phi_X$ with $|\lambda_A|^2(1-q)<1$, and $K=[K^{(1)},\ldots,K^{(\mathsf{I_o}-1)}]$ with
each matrix $K^{(i)}$ having compatible dimensions. It can be easily shown that $\mathcal{L}_K(X)$ is linear and non-decreasing on
the positive semi-definite cone.

The following result holds.


\begin{theorem}\label{thm:Contraction2LMI}
Suppose $|\lambda_A|^2(1-q)<1$. The following statements are equivalent:
\begin{enumerate}
\item[(i).] There {exists} $\,K\triangleq[K^{(1)},\ldots,K^{(\mathsf{I_o}-1)}]$ with each matrix
$K^{(i)}$ having compatible dimensions such that $\lim_{k\rightarrow \infty}\mathcal{L}^k_K(X)= 0$ for any
$X\in \mathbb{S}_+^n$;
\item[(ii).] There {exists}  $\,K\triangleq[K^{(1)},\ldots,K^{(\mathsf{I_o}-1)}]$ with each matrix
$K^{(i)}$ having compatible dimensions such that $|\lambda_{H(K)}|<1$;
\item[(iii).] There {exist}  $\,K\triangleq[K^{(1)},\ldots,K^{(\mathsf{I_o}-1)}]$ with each matrix
$K^{(i)}$ having compatible dimensions and  $P>0$ such that
$\mathcal{L}_K(P)<P$;
\item[(iv).] There exist $F_1,\ldots,F_{\mathsf{I_o}-1}$, $X>0,Y>0$ such that
\begin{equation}\label{eqn:LMI-2}
\left[\begin{array}{ccccc}
Y&\sqrt{1-q}A'Y&\sqrt{q}A'X\\
\sqrt{1-q}YA&Y&0\\
\sqrt{q}XA&0&X
\end{array}\right]\geq 0
\end{equation}
and
\begin{equation}\label{eqn:LMI-1}
\left[\begin{array}{cccccccccc}
X& \sqrt{p}(A'Y+C'F_1) & \cdots &\sqrt{p(1-p)^{\mathsf{I_o}-2}}\left((A^{\mathsf{I_o}-1})'Y+(C^{(i)})'F_{\mathsf{I_o}-1}\right)\\
\bf{*} & Y &\cdots & 0\\
\vdots & \vdots &\ddots & \vdots\\
* &
\bf{\ast} &\cdots & Y
\end{array}\right]>0,
\end{equation}
where $*$'s represent entries that are Hermitian conjugate of the
entries above the diagonal.
\end{enumerate}
If any of the above  statements holds,
then $\sup_{j\geq 1}\mathbb{E}\|P_{\beta_j}\|<\infty$, i.e., the Kalman filtering
system is peak-covariance stable.
\end{theorem}
The proof of Theorem~\ref{thm:Contraction2LMI} is given in Appendix C.
To summarize, Theorem~\ref{thm:Contraction2LMI} makes it possible to check the sufficient condition of Theorem \ref{thm:main-thm}
through an LMI feasibility criterion. It can be expected that,
given the ability to  search for $K^{(i)}$'s on a positive semi-definite cone, Theorem~\ref{thm:Contraction2LMI}
gives a less conservative condition than Proposition~\ref{corollary:main-result} does; this
 is demonstrated by Example I in Section~\ref{section:simulation}.

\begin{remark}
In~\cite{huang-dey-stability-kf,huang2006cdc,xie2008stability},
the criteria for peak-covariance stability are difficult to check since some constants related to the operator $g$ are hard to explicitly compute.
A thorough numerical search may be computationally demanding.
In contrast, the stability check of Theorem~\ref{thm:Contraction2LMI} uses an LMI feasibility problem, which can often be efficiently solved.
\end{remark}

\section{Mean-Square Stability}
\label{section:mss-KF}
In this section, we will discuss mean-square stability of
Kalman filtering with Markovian packet losses.
\begin{definition}\label{def:mms}
The Kalman filtering system
with packet losses is mean-square stable if $\sup_{k\in\mathbb{N}} \mathbb{E} \| P_{k}\|<\infty$.
\end{definition}
\subsection{From Peak-Covariance Stability to Mean-Square Stability}
Note that the peak-covariance stability characterizes the filtering system at stopping times defined by~\eqref{def:stopping-times}, while
mean-square stability characterizes the property of stability
at all sampling times. In the literature, the relationship between the
two stability notations is still an open problem. In this section, we aim to establish a
connection between peak-covariance stability and mean-square stability. Firstly, we need the following definition for the defective eigenvalues of a matrix.

\begin{definition}
For $\lambda\in\sigma(A)$ where $A$ is a matrix, if the algebraic multiplicity and
the geometric multiplicity of $\lambda$ are equal, then $\lambda$
is called a semi-simple eigenvalue of $A$. If $\lambda$ is
not semi-simple, $\lambda$ is called a defective eigenvalue of $A$.
\end{definition}

We are now able to present the following theorem indicating that as long as $A$ has no defective eigenvalues on the unit circle, i.e., the corresponding Jordan
block is $1\times 1$,
 peak-covariance stability always
implies mean-square stability. In fact, we are going to prove this connection for {\it general} random packet drop processes $\{ \gamma_k\}_{k\in \mathbb{N}}$, instead of
limiting to the Gilbert-Elliott model.

\begin{theorem}\label{thm:peak-2-usual-stability}
Let $\{ \gamma_k\}_{k\in \mathbb{N}}$ be a random process over an underlying probability space  $(\mathscr{S},\mathcal{S},\mu)$ with each $\gamma_k$ taking its value in
$\{0,1\}$. Suppose
$\{\beta_j\}_{j\in\mathbb{N}}$ take finite values $\mu-$almost surely,
and that $A$ has no defective eigenvalues on the unit circle. Then the peak-covariance stability of the Kalman filter always implies mean-square stability, i.e.,
$\sup_{k\in\mathbb{N}}\mathbb{E}\|P_{k}\|<\infty$ whenever  $\sup_{j\in\mathbb{N}}\mathbb{E}\|P_{\beta_j}\|<\infty$.
\end{theorem}
Note that $\{\beta_j\}_{j\in\mathbb{N}}$ can be defined over any random packet loss processes, therefore the peak-covariance stability with packet losses that the filtering system is undergoing remains in accord with  Definition~\ref{def:peak-cov-stability}.

Theorem \ref{thm:peak-2-usual-stability} bridges the two stability notions of Kalman filtering with random packet losses in the literature. Particularly this connection covers most of the existing models for packet losses, e.g., i.i.d. model~\cite{Sinopoli2004}, bounded Markovian~\cite{xiao2009kalman}, Gilbert-Elliott~\cite{huang2006cdc}, and finite-state channel~\cite{wang1995finite,zhang1999finite} Although $\sup_{k\in\mathbb{N}}\mathbb{E}\|P_k\|$ and $\sup_{j\in\mathbb{N}}\mathbb{E}\|P_{\beta_j}\|$
are not equal in general,
this connection is built upon a critical understanding that, no matter to which inter-arrival interval between two successive $\beta_j$'s
the time $k$ belongs, \textit{$\|P_k\|$ is uniformly bounded from above by an affine function of the norm of the
peak covariances at the starting and ending points thereof.}
This point holds regardless of
the model of packet loss process. The proof of Theorem \ref{thm:peak-2-usual-stability} was given in Appendix D.

\begin{figure}[t!]
\begin{center}
\includegraphics[width=3.75in]{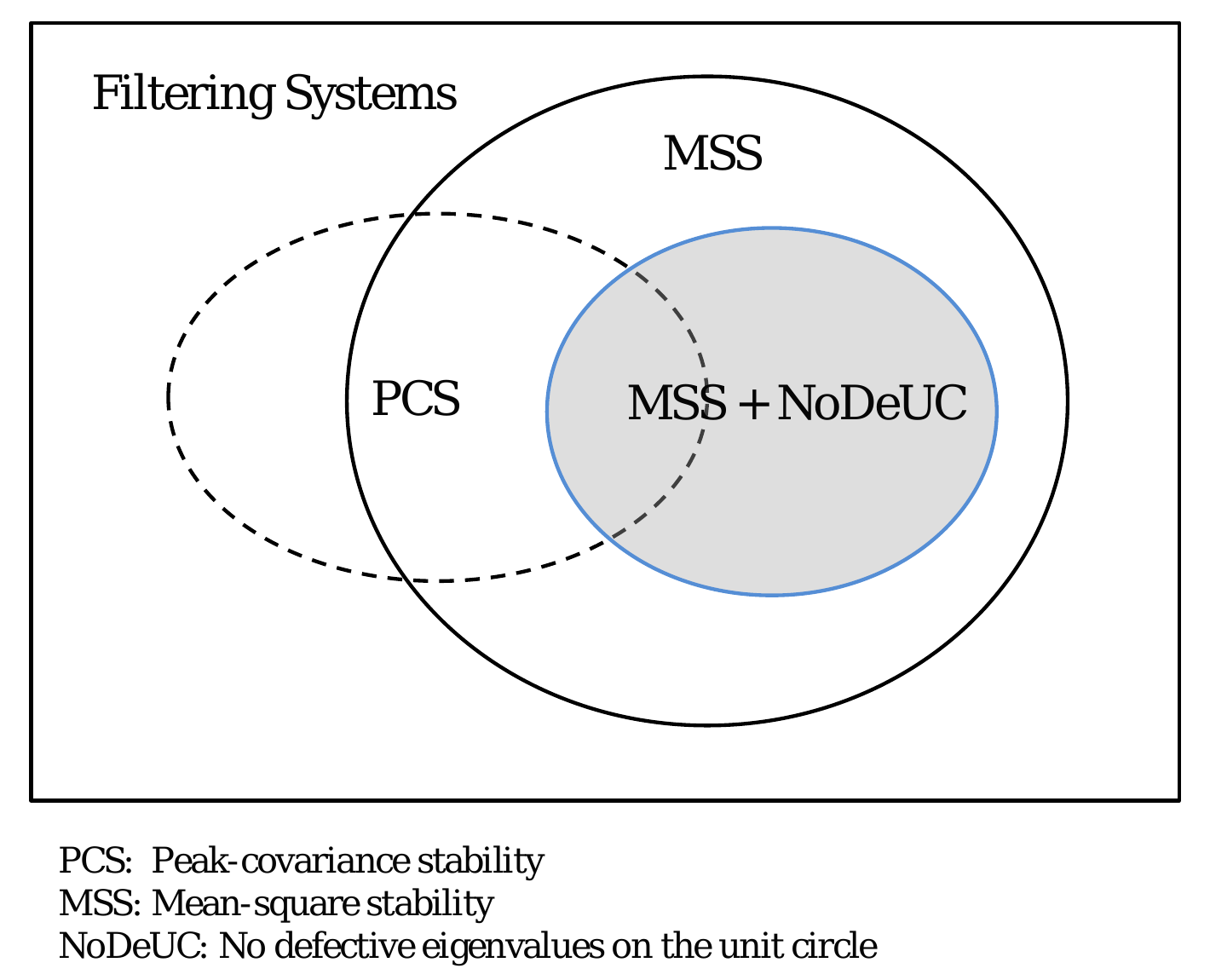}
\caption{Relationships between the peak-covariance stability and mean-square stability over the space of the filtering systems under consideration. Theorem \ref{thm:peak-2-usual-stability} indicates that the intersection of the two sets,  PCS and MSS+NoDEUC, is contained in the set MSS.  }\label{fig:relationship-mms-pcs}
\end{center}
\end{figure}

We also remark that there is some difficulty in relaxing the assumption that $A$ has no defective eigenvalues on the unit circle  in  Theorem \ref{thm:peak-2-usual-stability}. This is due to the fact that $A$'s defective eigenvalues on the unit circle will influence both the peak-covariance stability and mean-square stability in a nontrivial manner. See Fig.~\ref{fig:relationship-mms-pcs}.
\begin{remark}
In~\cite{huang-dey-stability-kf}, for a scalar model with i.i.d. packet losses,
it has been shown that
the peak-covariance stability is equivalent to mean-square stability, while for
a vector system even with i.i.d. packet losses,
the relationship between the two is  unclear. In~\cite{xie2008stability},
the equivalence between the two stability notions was established for
systems that are one-step observable, again for the i.i.d. case.
Theorem~\ref{thm:peak-2-usual-stability} now fills the gap for a large class of vector systems under general random packet drops.
\end{remark}

\subsection{The Critical $p-q$ Curve}%
In this subsection,  we first show that for a fixed $q$ in the Gilbert-Elliott channel,
there exists a critical failure rate $p_c$, such that
if and only if the failure rate is below $p_c$, the Kalman filtering is mean-square stable.
This conclusion is relatively independent of previous results, and the proof
relies on a coupling argument and can be found in Appendix E.
\begin{proposition}\label{thm:critical-value-p}
Let the recovery rate $q$ satisfy $|\lambda_A|^2(1-q)<1$. Then there exists
a critical value $p_c\in(0,1]$ for the failure rate in the sense that \begin{itemize}
\item[(i)] $\sup_{k\in\mathbb{N}}\mathbb{E}\|P_k\|<\infty$ for all $\Sigma_0\geq0$ and $0<p<p_c~$;
\item[(ii)] there exists $ \Sigma_0\geq0$ such that $\sup_{k\in\mathbb{N}}\mathbb{E}\|P_k\|=\infty$ for all $p_c< p<1$.
\end{itemize}
\end{proposition}

\begin{figure}[t!]
\begin{center}
\includegraphics[width=4.0in]{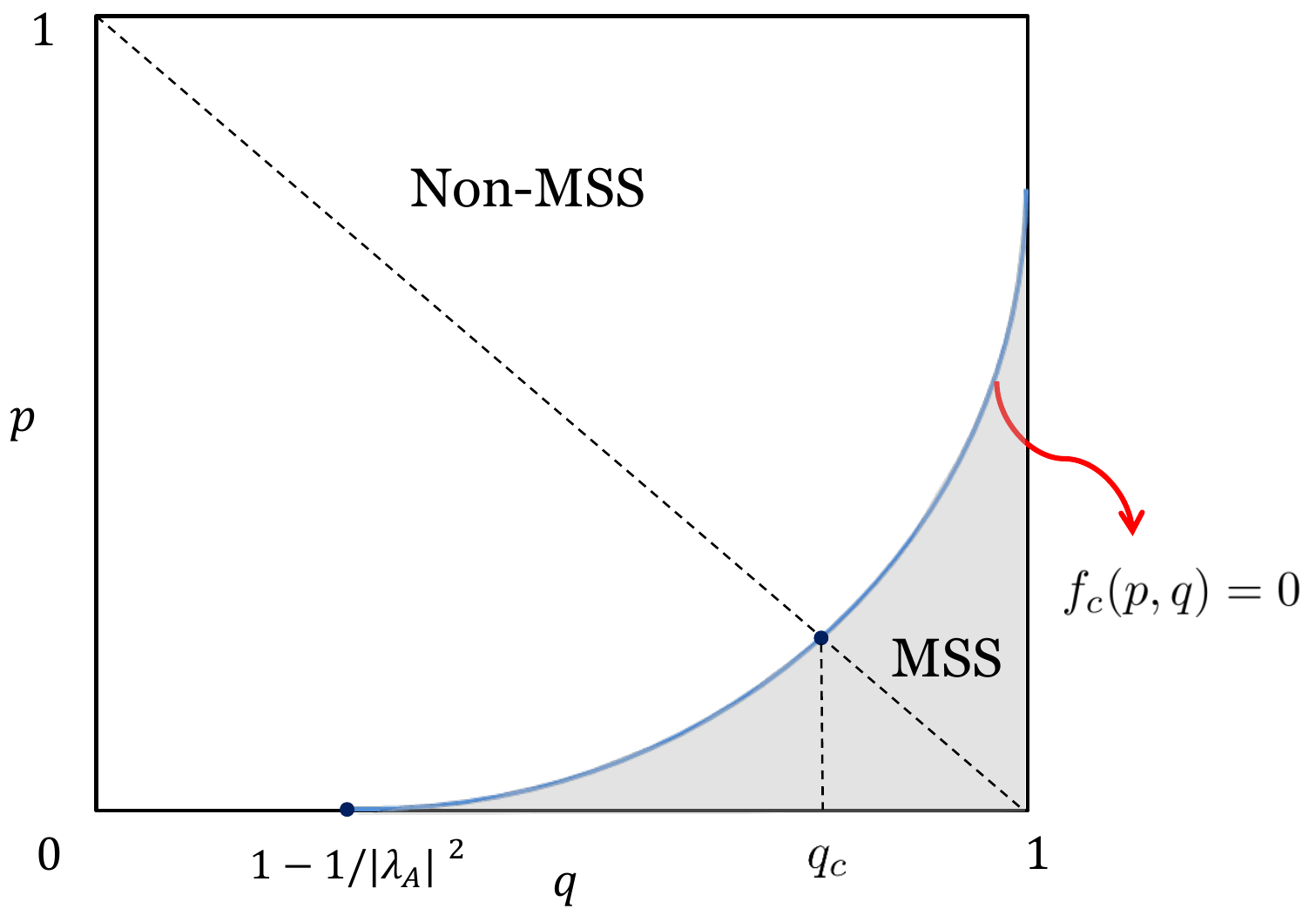}
\caption{The $p-q$ plane is divided into MSS (Mean-Square Stability) and Non-MSS regions by the  critical curve $f_c(p,q)=0$. When $p+q=1$, the Markovian packet loss process is reduced to an  i.i.d. process.  As a result, the intersection point of the curves $f_c(p,q)=0$ and $p+q=1$ gives the critical packet drop probability established in \cite{sinopoli}. }\label{fig:p-q-critical-curve}
\end{center}
\end{figure}

It has been shown in~\cite{you2011mean} that a
necessary condition for mean-square stability of the filtering system
is $|\lambda_{A}|^2(1-q)< 1$, which is only related to the recovery rate $q$.
For Gilbert-Elliot channels, a critical value
phenomenon with respect to $q$ is also expectable. Theorem~\ref{proposition:critical-curve} proves the existence of the critical $p-q$ curve and Fig.~\ref{fig:p-q-critical-curve} illustrates this critical curve in the $p-q$ plane.
The proof,  analogous to that of Proposition~\ref{thm:critical-value-p},
is given in Appendix~F.
\begin{theorem}\label{proposition:critical-curve}
There exists
a critical curve defined by $f_c(p,q)=0$,  which reads two  non-decreasing functions $p=p_c(q)$ and $q=q_c(p)$ with $q_c(\cdot)=p_c^{-1}(\cdot)$,
dividing $(0,1)^2$ into two disjoint regions such that:
\begin{itemize}
\item[(i)] If $(p,q)\in \big\{f_c(p,q)>0\big\}$, then $\sup_{k\in\mathbb{N}}\mathbb{E}\|P_k\|<\infty$ for
all $\Sigma_0\geq0$;
\item[(ii)]  If $(p,q)\in \big\{f_c(p,q)<0\big\}$, then there exists $ \Sigma_0\geq0$ under which
$\sup_{k\in\mathbb{N}}\mathbb{E}\|P_k\|=\infty$.
\end{itemize}

\end{theorem}

\begin{remark}
If the packet loss process is an i.i.d. process, where $p+q=1$ in the
transition probability matrix defined in~\eqref{eqn:markov-trnsition-prob},
Proposition~\ref{thm:critical-value-p} and Theorem~\ref{proposition:critical-curve}
recover the result of Theorem 2 in~\cite{Sinopoli2004}. It is worth pointing out that whether mean-square stability holds or not exactly on the curve $f_c(p,q)=0$ is beyond the reach of the current analysis (even for the i.i.d. case with $p+q=1$): such an understanding relies on the  compactness of the stability or non-stability regions.
\end{remark}

\subsection{Mean-square Stability Conditions}
We can now make use of the peak-covariance stability conditions we obtained in the last section, and the connection between peak-covariance stability and mean-square stability indicated in Theorem~\ref{thm:peak-2-usual-stability}, to establish mean-square stability conditions for the considered Kalman filter. It turns out that the assumption requiring  no defective eigenvalues on the unit circle, can be relaxed by an approximation method. We present the following result.

\begin{theorem}\label{thm:sufficient-cond}
Let the recovery rate $q$ satisfy $|\lambda_A|^2(1-q)<1$. Then
there holds  $p_c\geq \underline{p}$, where
\begin{equation}\label{eqn:upper-bound-p}
\underline{p}\triangleq \sup\Big\{p:\exists (K,P)\mathrm{~s.t.~}\mathcal{L}_K(P)<P,P>0\Big\},
\end{equation}
i.e., for all $\Sigma_0\geq 0$ and $0<p<\underline{p}$, the Kalman filtering system is mean-square stable.
\end{theorem}
The proof of Theorem \ref{thm:sufficient-cond} was given in Appendix G.
\begin{remark}
For second-order systems and certain classes of high-order systems, such as
non-degenerate systems, necessary and sufficient conditions for mean-square stability have been
derived in~\cite{you2011mean} and~\cite{yilin12criticalvalue}.
However, these results rely on a particular system structure and fail to
apply to general LTI systems. It seems challenging to find
an explicit description of necessary and sufficient conditions
for mean-square stability of general LTI systems.  Theorem~\ref{thm:sufficient-cond}
gives a stability criterion
for general LTI systems.
\end{remark}

\section{Numerical Examples}\label{section:simulation}
In this section, we present  two numerical  examples to
demonstrate the theoretical results we established  in Sections~\ref{section:peak-covariance-Stability} and~\ref{section:mss-KF}.

\subsection{Example I: A Second-order System}
To compare with the works in~\cite{huang2006cdc,huang-dey-stability-kf,xie2007peak}, we will
examine the same vector example considered therein. The parameters are specified as follows:
$$
A=\left[\begin{array}{ccc}
1.3 & 0.3 \\
0 & 1.2
\end{array}\right],~~
C=[1,1],
$$
$Q=I_{2\times 2}$ and $R=1$. As illustrated in~\cite{huang-dey-stability-kf}, it is easily checked that
$\mathsf{I_o}=2$ and the spectrum of $A$ is $\sigma(A)=\{1.2,1.3\}$, and that
$\lambda_A=1.3$.

First let us compare the sufficient condition we provide in Proposition~\ref{corollary:main-result} with the counterpart provided in~\cite{huang-dey-stability-kf}.
Note that  $|\lambda_A|^2(1-q)<1$ is a necessary condition for mean-square stability.
We take $q=0.65$ as was done in~\cite{huang-dey-stability-kf}.
As for the failure rate $p$,~\cite{huang-dey-stability-kf} concludes that
$p<0.04$ guarantees peak-covariance stability; while Proposition~\ref{corollary:main-result} requires
$$
p<\frac{1-|\lambda_A|^2(1-q)}{|\lambda_A|^4q},
$$
which generates the less conservative condition $p<0.22$.
In~\cite{huang-dey-stability-kf}, all numerical simulations were implemented with
parameters $(p,q)=(0.03,0.65)$.
Note that, for the associated channel with $(p,q)=(0.04,0.65)$,
$\mathbb{P}(\gamma_k=0)=0.0580$ when the packet loss process enters the stationary
distribution, which means that the allowed long term packet loss rate is at most
$5.80\%$. However, by choosing a larger $p$, Proposition~\ref{corollary:main-result} permits $\mathbb{P}(\gamma_k=0)=0.2529$ at the stationary distribution at most, i.e., the allowed  long term packet dropout rate is $25.29\%$.
Similarly, the example in~\cite{xie2007peak} allows $p=0.1191$ at most.
Separately, we note that it is rather convenient to check the condition in Proposition~\ref{corollary:main-result} even with manual calculation; in contrast, to check the conditions in~\cite{huang-dey-stability-kf} and~\cite{xie2007peak} involves a considerable amount of numerical calculation.

\begin{figure}[t!]
\begin{center}
\includegraphics[width=5in]{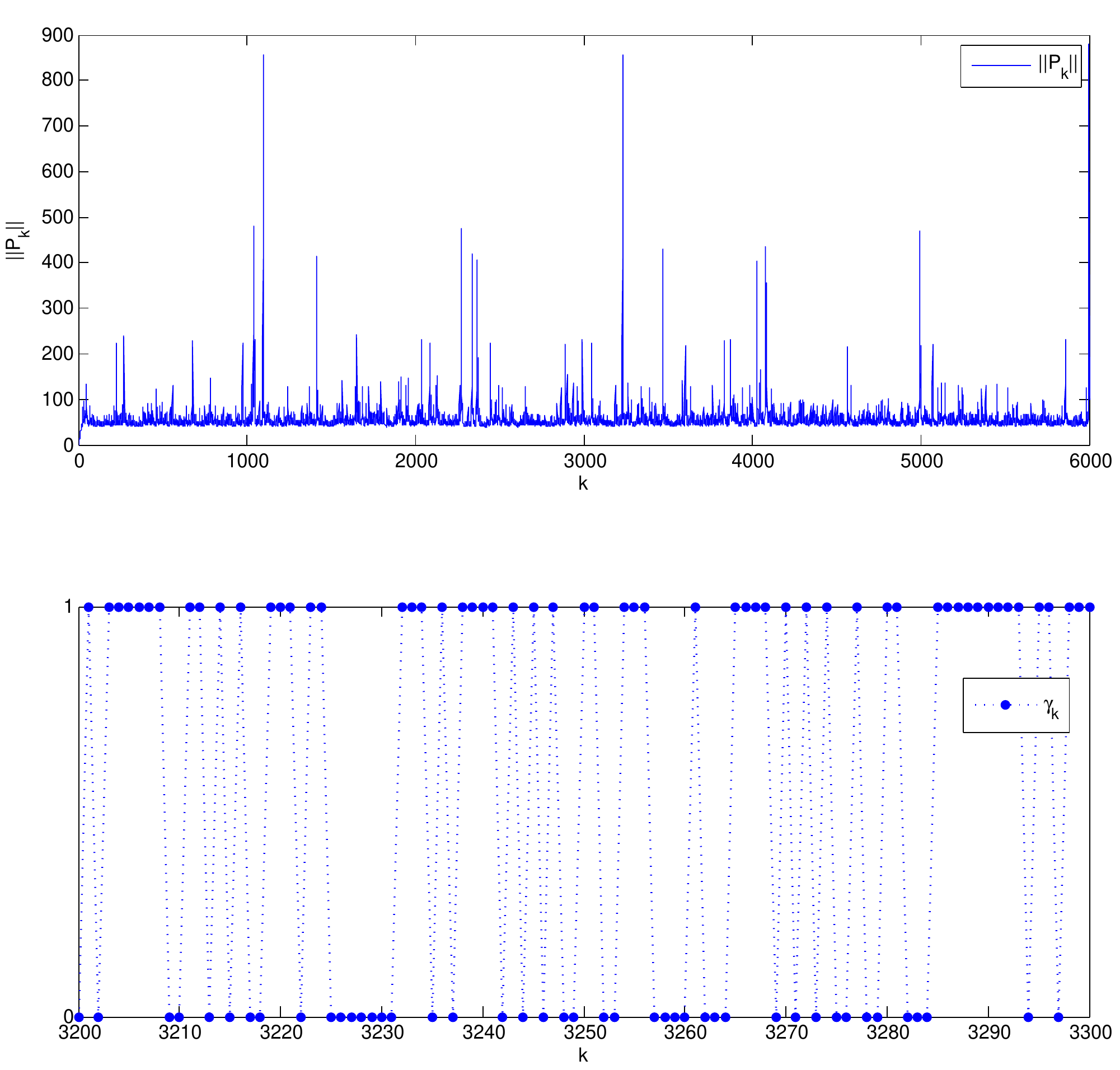}
\caption{A sample path of $\|P_k\|$ and $\gamma_k$ with $p=0.5$, $q=0.65$ in
Example I.}\label{sample_path_Pk_p_050}
\end{center}
\end{figure}

\begin{figure}[t!]
\begin{center}
\includegraphics[width=5in]{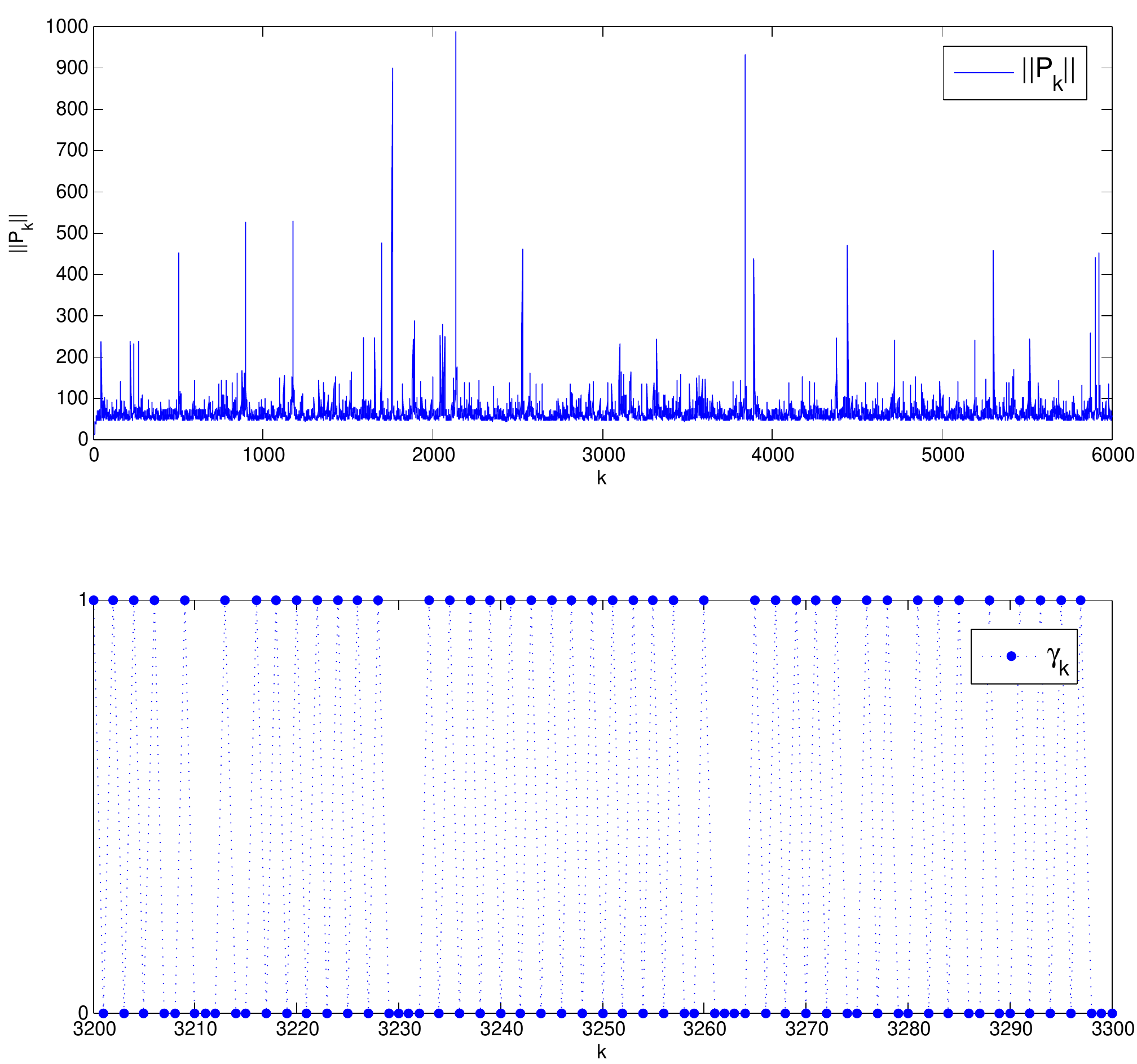}
\caption{A sample path of $\|P_k\|$ and $\gamma_k$ with $p=0.99$, $q=0.65$ in
Example I.}\label{sample_path_Pk_p_099}
\end{center}
\end{figure}

In what follows, we use
the criterion established in Theorem~\ref{thm:Contraction2LMI} to check for the peak-covariance stability.
Then we obtain that when $p=1$ the LMI in 2) of Theorem~\ref{thm:Contraction2LMI}
is still feasible.\footnote{\label{footnote:1}
To satisfy the assumption~\ref{asmpt:assumpt-A2}, we need to configure
$p=1-\epsilon$ for an arbitrary small positive $\epsilon$.}
It should be pointed that at least for the parameters specified in this example, the criterion of~\cite{huang-dey-stability-kf}
only covers the Gilbert-Elliott models with failure rate lower than $4.5\%$.
Fig.~\ref{sample_path_Pk_p_050} and Fig.~\ref{sample_path_Pk_p_099} illustrate
sample paths of $\|P_k\|$ and $\gamma_k$ with $(p,q)=(0.5,0.65)$ and $(p,q)=(0.99,0.65)$, respectively. The figures show that even a high value of $p$ may not effect the peak-covariance
stability with the system parameters specified in this example, showing that Theorem~\ref{thm:Contraction2LMI} provides
a less conservative
criterion than Proposition~\ref{corollary:main-result} or~\cite{huang-dey-stability-kf,xie2007peak} does, a fact which is consistent with
the theoretical analysis in Section~\ref{section:peak-covariance-Stability}.

\subsection{Example II:~A Third-order System}
\begin{figure}[t!]
\begin{center}
\includegraphics[width=5.3in]{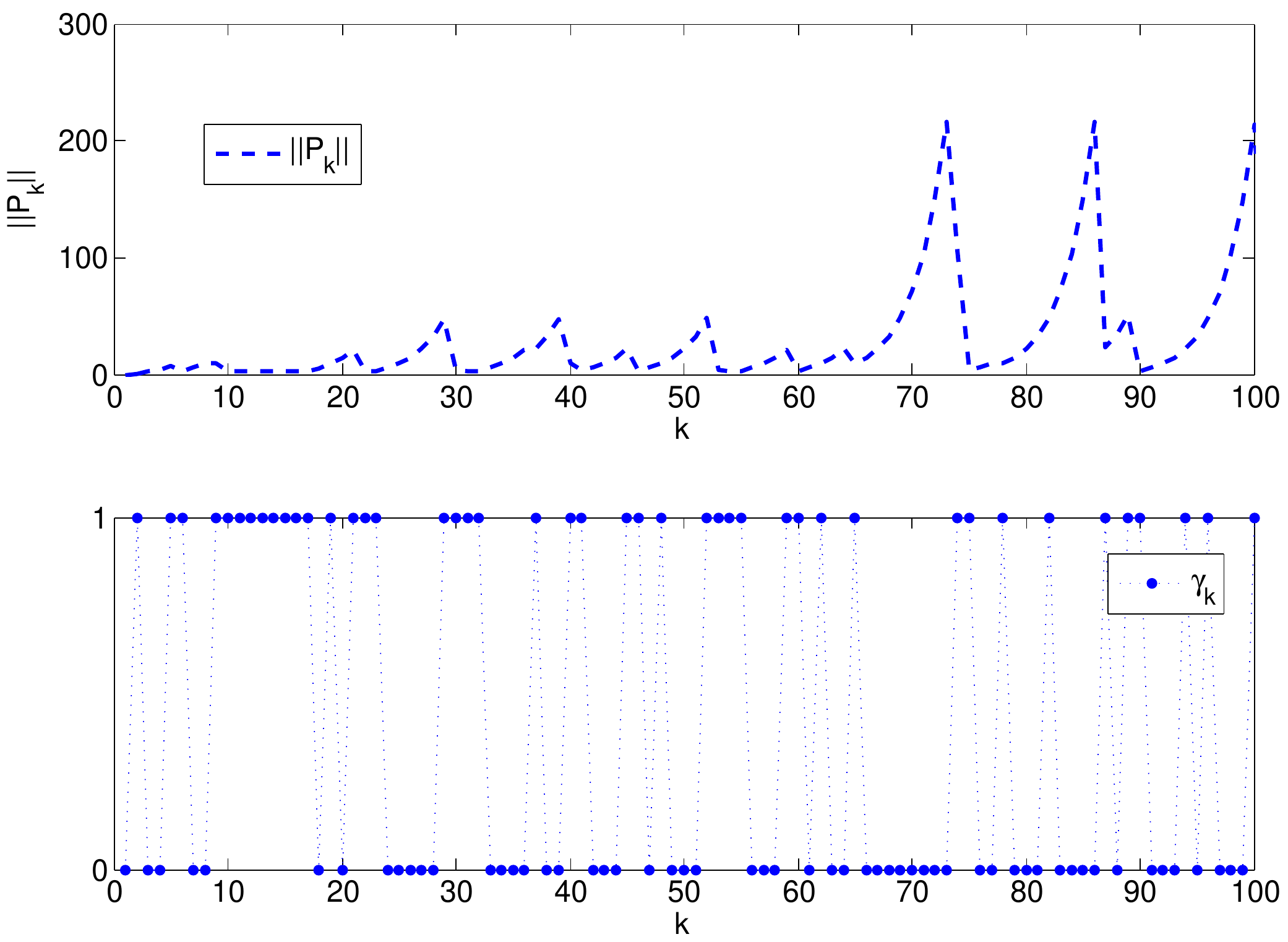}
\caption{A sample path of $\|P_k\|$ and $\gamma_k$ with $p=0.45$, $q=0.5$ in
Example II.}\label{sample_path_Pk_example2_bounded}
\end{center}
\end{figure}

\begin{figure}[t!]
\begin{center}
\includegraphics[width=5.1in]{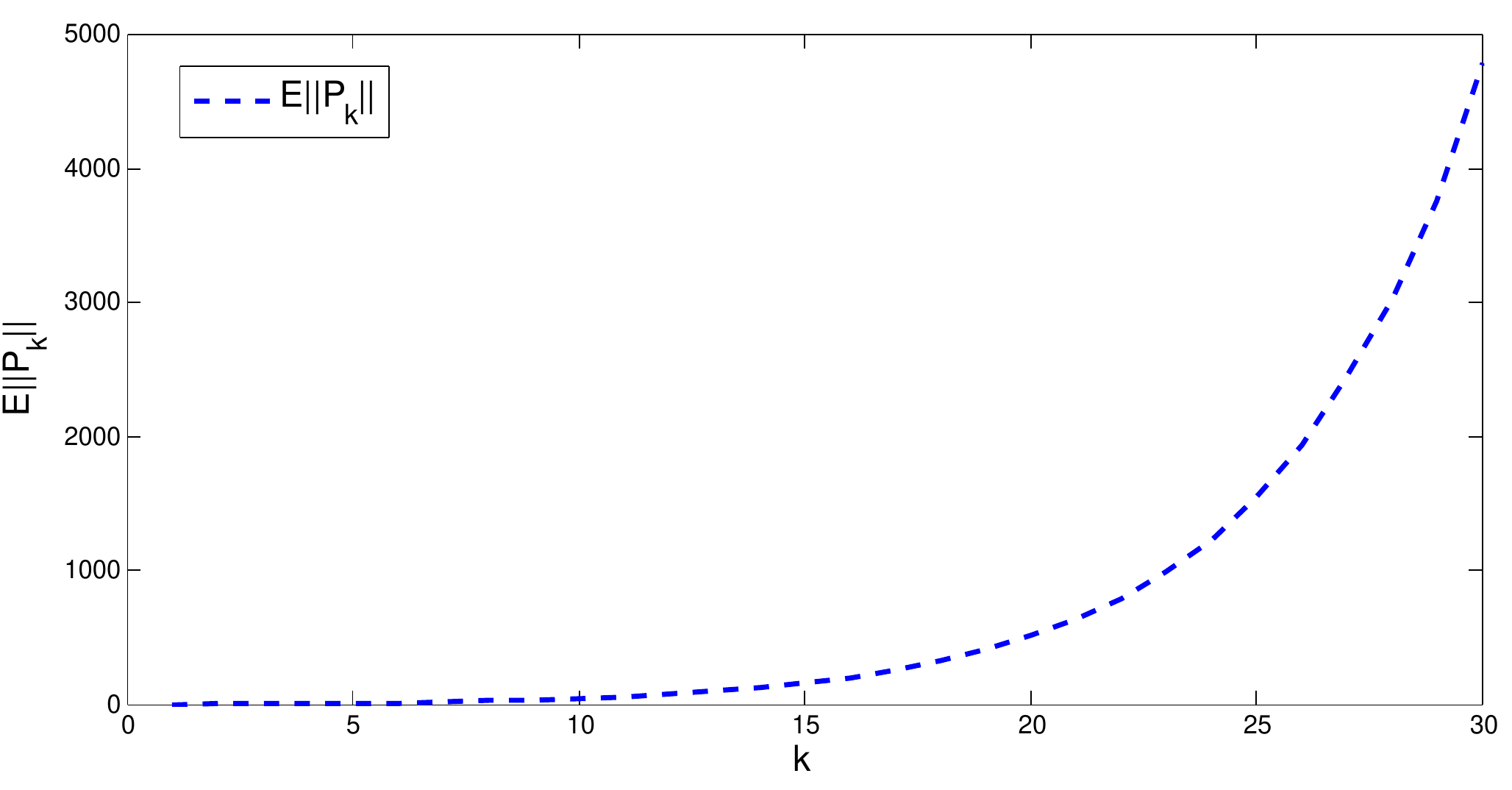}
\caption{Divergence of $\mathbb{E}\|P_k\|$ with $p=0.99$, $q=0.5$ in
Example II.}\label{sample_path_Pk_example2_unbounded}
\end{center}
\end{figure}

To compare the work in Section~\ref{section:mss-KF} with the result~\cite{you2011mean} and~\cite{yilin12criticalvalue}, we will use the following
example, where the parameters are given by
\begin{equation}\label{eqn:exmaple2}
A=\left[\begin{array}{ccc}
1.2 & 0 & 0 \\
0 & 1.2 & 0 \\
0 & 0 & -1.2
\end{array}\right],~~
C=\left[\begin{array}{ccc}
1 &0 &1\\
0 &1 &1\end{array}\right],
\end{equation}
$Q=I_{3\times 3}$ and $R=I_{2 \times 2}$.
In~\cite{you2011mean}
and~\cite{yilin12criticalvalue}, mean-square stability of Kalman filtering for so-called
non-degenerate systems has been studied. Before proceeding, we introduce their definition, which originates from~\cite{yilin10bound}. \begin{definition}
Consider a system $(C,A)$ in diagonal
standard form, i.e., $A=\mathrm{diag}
(\lambda_1,\ldots,\lambda_n)$ and
$C=[C_1,\ldots,C_n]$. A quasi-equiblock
of the system is defined as a subsystem
$(C_\mathcal{I},A_\mathcal{I})$, where
$\mathcal{I}\triangleq \{l_1,\ldots,l_i\}
\subset\{1.\ldots,n\}$, such that
$A_\mathcal{I}=\mathrm{diag}(\lambda_{l_1},\ldots,\lambda_{l_i})$
with $|\lambda_{l_1}|=\cdots=|\lambda_{l_i}|$ and
$C_\mathcal{I}=[C_{l_1},\ldots,C_{l_i}]$.
\end{definition}
\begin{definition}
A diagonalizable system $(C,A)$ is non-degenerate if
every quasi-equiblock of the system is one-step observable. Conversely, it is
degenerate if it has at least one quasi-equiblock that is
not one-step observable.
\end{definition}
By the above definitions, the system in~\eqref{eqn:exmaple2}
is observable but degenerate since $|\lambda_1|=|\lambda_2|=|\lambda_3|$
but $(C,A)$ is not one-step observable.
Hence none of the necessary and sufficient conditions developed
in the aforementioned two papers is applicable in this example.
To the best of our knowledge, no tool has been established so far
to study mean-square stability of such a system.
The results presented in
Section~\ref{section:mss-KF} provide us a universal criterion for mean-square stability. Let us fix $q=0.5.$
We can conclude from Theorem~\ref{thm:sufficient-cond} that
if $p\leq 0.465$ the Kalman filter is mean-square stable.
Fig.~\ref{sample_path_Pk_example2_bounded} illustrates a
sample path of $\|P_k\|$ and $\gamma_k$ with $(p,q)=(0.45,0.5)$.
Fig.~\ref{sample_path_Pk_example2_unbounded} illustrates that with $(p,q)=(0.99,0.5)$
the expected
prediction error covariance matrices diverge.
One can verify that
when $q=0.5$ and $p=1$ the criterion in Theorem~\ref{thm:sufficient-cond}
is violated as the LMI in Theorem~\ref{thm:Contraction2LMI} is infeasible.


\section{Conclusions}\label{section:conclusions}
We have investigated  the stability of Kalman filtering
over Gilbert-Elliott channels. Random packet drop follows a time-homogeneous two-state Markov chain where the two states indicate successful or failed packet transmissions.
We established  a relaxed condition guaranteeing   peak-covariance stability
described by an inequality in terms of the spectral radius
of the system matrix and transition probabilities of
the Markov chain, and then showed that
 the condition can be reduced to an LMI feasibility problem. It was proved that peak-covariance stability implies mean-square stability if the system matrix has no defective eigenvalues on the
unit circle. This connection holds for general random packet drop processes.  We also proved that there exists a critical region
in the $p-q$ plane such that
if and only if the pair of recovery and failure rates
falls into that region the expected prediction error covariance matrices
are uniformly bounded.
By fixing the recovery rate,
a lower bound for the critical failure rate was obtained  making use of  the relationship between two stability criteria for  general
LTI systems.
Numerical examples
demonstrated  significant improvement on the  effectiveness of our appraoch compared  with the existing literature.

\vspace{20mm}

\noindent{\Large \bf Appendices. Proofs of Statements}

\section*{Appendix A. Auxiliary Lemmas}
In this section, we collect some lemmas that are regularly  used throughout the proofs
of our main results.

\begin{lemma}[Lemma A.1 in~\cite{shi-tac10}]\label{lemma:appendix-g-h}
For any matrices $X\geq Y\geq 0$, the following inequalities hold
\begin{eqnarray}
h(X)&\geq& h(Y),\label{eqn:h-monotone}\\
g(X)&\geq& g(Y),\label{eqn:g-monotone}\\
h(X)&\geq& g(X),\label{eqn:h-geq-g}
\end{eqnarray}
where the operators $h$ and $g$ are defined in~\eqref{eqn:h-func} and~\eqref{eqn:g-func} respectively.
\end{lemma}

\begin{lemma}\label{lemma:gx}
Consider the operator
\begin{eqnarray*}\phi_i({K}^{(i)},P)&=&(A^i+K^{(i)}C^{(i)})X(\cdot)^*+
[A^{(i)}~K^{(i)}]\left[\begin{array}{ccc}
Q^{(i)}& Q^{(i)}(D^{(i)})'\\
\ast& D^{(i)}(Q^{(i)})(D^{(i)})'+R^{(i)}
\end{array}\right][A^{(i)}~K^{(i)}]^*,\\
&&~\forall i\in \mathbb{N},
\end{eqnarray*}
where $C^{(i)}=[C',~A'C',\cdots, (A')^{i-1}C']'$, $A^{(i)}=[A^{i-1},\cdots,A,~I]$,
$D^{(i)}=0$ for $i=1$ otherwise $D^{(i)}=\left[\begin{array}{cccc}
0& 0 & \cdots & 0\\
C& 0 & \cdots & 0\\
\vdots &\vdots& \ddots& \vdots\\
CA^{i-2}& CA^{i-3}& \cdots & 0\end{array}\right]$,
$Q^{(i)}=\mathrm{diag}(\underbrace{Q,\cdots,Q}_i)$,
$R^{(i)}=\mathrm{diag}(\underbrace{R,\cdots,R}_i)$,
and $K^{(i)}$ are of compatible dimensions.
 For any $X\geq 0$ and $K^{(i)}$,
it always holds that
$$g^i(X)=\min_{K^{(i)}}\phi_i(K^{(i)},X)\leq \phi_i(K^{(i)},X).$$
\end{lemma}
{\it Proof.}
The result is readily established
when setting $B=I$ in Lemmas 2 and 3 in~\cite{xiao2009kalman}.
For $i=1$, The result is well known as Lemma 1 in~\cite{Sinopoli2004}.
\hfill$\square$


\begin{lemma}[\cite{solo91}]\label{lemma:norm2radius}
For any $A\in\mathbb{C}^{n\times n}$, $\epsilon>0$ and $k\in\mathbb{N}$, it holds that
$$\|A^k\|\leq \sqrt{n}(1+2/\epsilon)^{n-1}\Big(|\lambda_A|+\epsilon\|A\|\Big)^k$$
\end{lemma}

\begin{lemma}[Lemma 2 in~\cite{Costa04TAC}]\label{lemma:matrix-depomositation}
For any $G\in \mathbb{C}^{n\times n}$ there exist $G_i\in
\mathbb{S}^{n}_+,~i=1,2,3,4$ such that
$$G=(G_1-G_2)+(G_3-G_4)\mathbf{i}$$
where $\mathbf{i}=\sqrt{-1}$.
\end{lemma}

\section*{Appendix B. Proof of Theorem~\ref{thm:main-thm}}
The proof relies on the following two lemmas.
\begin{lemma}[Lemma 5 in~\cite{huang-dey-stability-kf}]\label{lemma:riccati>Io}
Assume that $(C,A)$ is observable and $(A,Q^{{1}/{2}})$ is controllable.
Define
$$
\mathbb{S}_0^n=\{P: 0\leq P\leq AP_0A'+Q,\hbox{~for some~} P_0\geq 0\},
$$
Then there exists a constant $\mathsf{L}>0$ such that
\begin{enumerate}
\item[(i).] for any $X \in \mathbb{S}_0^n$, $g^k(X)\leq \mathsf{L}I$ for all $k\geq \mathsf{I_o}$;
\item[(ii).] for any $X \in \mathbb{S}_+^n$, $g^{k+1}(X)\leq \mathsf{L}I$ for all $k\geq \mathsf{I_o}$,
\end{enumerate}
where the operator $g$ is defined in~\eqref{eqn:g-func}.
\end{lemma}

\begin{lemma}\label{lemma:matrix-series-convergence}
For $q\in(0,1)$ and $A\in\mathbb{R}^{n\times n}$, the series of matrices $\sum_{i=1}^\infty (A\otimes A)^i(1-q)^{i-1}q$
and $\sum_{i=1}^\infty \sum_{j=0}^{i-1}(A\otimes A)^j(1-q)^{i-1}q$ converge if and only if $|\lambda_A|^2(1-q)<1$.
\end{lemma}
{\it Proof.}
First observe that
$$\sum_{i=1}^\infty \sum_{j=0}^{i-1}(A\otimes A)^j(1-q)^{i-1}q
=\sum_{i=0}^\infty (A\otimes A)^i(1-q)^{i}.$$
The geometric series generated by $(A\otimes A)(1-q)$ converges if and only if
$|\lambda_{A\otimes A}|(1-q)<1$. Therefore the conclusion follows from
the fact that $|\lambda_{A\otimes A}|=\max\{|\lambda_i\lambda_j|:\lambda_i,\lambda_j\in \sigma(A)\}=|\lambda_A|^2$.
\hfill$\square$

Now fix  $j\geq 1$.
First note that, for any $k\in [\tau_{j+1},\beta_{j+1}-1]$, $\gamma_k=0$.
Hence we have
\begin{eqnarray}\label{eqn:peak-Pk-1}
P_{\beta_{j+1}}&=&\sum_{i=1}^{\infty}1_{\{\beta_{j+1}^*=i\}}
h^i(P_{\tau_{j+1}})\notag\\
&\triangleq&\sum_{i=1}^{\infty}1_{\{\beta_{j+1}^*=i\}}
A^{i}P_{\tau_{j+1}}(A^i)'+
\sum_{i=1}^{\infty}1_{\{\beta_{j+1}^*=i\}}V_i,
\end{eqnarray}
where $V_i\triangleq\sum_{l=0}^{i-1}A^lQ(A^l)'$.
Now let us consider the interval $[\beta_j,\tau_{j+1}-1]$ over which
$\tau_{j+1}^*$ packets are successfully received. We will analyze the relationship
between $P_{\tau_{j+1}}$ and $P_{\beta_j}$ in two separated cases, which are $\tau_{j+1}^*\leq \mathsf{I_o}-1$ and $\tau_{j+1}^*\geq \mathsf{I_o}$. Computation yields the following result
\begin{eqnarray}\label{eqn:peak-Pk-2}
P_{\tau_{j+1}}&=&\sum_{i=1}^{\mathsf{I_o}-1}
1_{\{\tau_{j+1}^*=i\}}g^{i}(P_{\beta_j})+\sum_{l=\mathsf{I_o}}^{\infty}
1_{\{\tau_{j+1}^*=l\}}g^{l}(P_{\beta_j})\notag\\
&\leq & \sum_{i=1}^{\mathsf{I_o}-1}
1_{\{\tau_{j+1}^*=i\}}g^{i}(P_{\beta_j})+\mathsf{L}I\sum_{l=\mathsf{I_o}}^{\infty}
1_{\{\tau_{j+1}^*=l\}}\notag\\
&\leq &\sum_{i=1}^{\mathsf{I_o}-1}
1_{\{\tau_{j+1}^*=i\}}\phi_i(K^{(i)},P_{\beta_j})+\mathsf{L}I\sum_{l=\mathsf{I_o}}^{\infty}
1_{\{\tau_{j+1}^*=l\}}\notag\\
&= &\sum_{i=1}^{\mathsf{I_o}-1}
1_{\{\tau_{j+1}^*=i\}}(A^i+K^{(i)}C^{(i)})P_{\beta_j}(A^i+K^{(i)}C^{(i)})^*
+\mathsf{L}I\sum_{j=\mathsf{I_o}}^{\infty}
1_{\{\tau_{j+1}^*=l\}}\notag\\
&&\hspace{3cm}+\sum_{i=1}^{\mathsf{I_o}-1}
1_{\{\tau_{j+1}^*=i\}}
[A^{i}~K^{(i)}]J_i[A^i~K^{(i)}]^*
\notag\\
&\triangleq& (A^i+K^{(i)}C^{(i)})P_{\beta_j}(A^i+K^{(i)}C^{(i)})^*+U,
\end{eqnarray}
where $J_i\triangleq\left[\begin{array}{ccc}
Q^{(i)}& Q^{(i)}(D^{(i)})'\\
D^{(i)}(Q^{(i)})& D^{(i)}(Q^{(i)})(D^{(i)})'+R^{(i)}
\end{array}\right]$ and $$U\triangleq \mathsf{L}I\sum_{j=\mathsf{I_o}}^{\infty}
1_{\{\tau_{j+1}^*=l\}}+\sum_{i=1}^{\mathsf{I_o}-1}
1_{\{\tau_{j+1}^*=i\}}
[A^{i}~K^{(i)}]J_i[A^{i}~K^{(i)}]^*$$ is bounded. The first inequality is from Lemma~\ref{lemma:riccati>Io} and the
second one follows from Lemma~\ref{lemma:gx}.
By substituting~\eqref{eqn:peak-Pk-2} into~\eqref{eqn:peak-Pk-1}, it yields
\begin{eqnarray}\label{eqn:peak-Pk-3}
P_{\beta_{j+1}}&\leq& \sum_{i=1}^\infty 1_{\{\beta_{j+1}^*=i\}}A^i\left[
\sum_{l=1}^{\mathsf{I_o}-1}1_{\{\tau_{j+1}^*=l\}}(A^l+K^{(l)}C^{(l)})P_{\beta_j}(A^l+K^{(l)}C^{(l)})^*
\right](A^i)'+W,
\end{eqnarray}
where $W\triangleq\sum_{i=1}^\infty 1_{\{\beta_{j+1}^*=i\}}A^iU(A^i)'+\sum_{i=1}^\infty 1_{\{\beta_{j+1}^*=i\}}V_i$.

To facilitate discussion, we force $P_{\beta_{j+1}}$ in~\eqref{eqn:peak-Pk-3} to take the maximum,
\begin{equation}\label{eqn:peak-Pk-4}
P_{\beta_{j+1}}=\sum_{i=1}^\infty 1_{\{\beta_{j+1}^*=i\}}A^i\left[
\sum_{l=1}^{\mathsf{I_o}-1}1_{\{\tau_{j+1}^*=l\}}(A^l+K^{(l)}C^{(l)})P_{\beta_j}(A^l+K^{(l)}C^{(l)})^*
\right](A^i)'+W.
\end{equation}
In what follow, we only take~\eqref{eqn:peak-Pk-4} into consideration.
For other cases in~\eqref{eqn:peak-Pk-3}, the subsequent conclusion still holds as
\eqref{eqn:peak-Pk-4} renders an upper envelop of $\{P_{\beta_{j}}\}_{j\in\mathbb{N}}$.

We introduce the  vectorization operator. Let
$X=[x_1~x_2~\cdots~x_n]\in\mathbb{C}^{m\times n}$ where $x_i\in\mathbb{C}^m$.
Then we define
$$
\mathrm{vec}(X)\triangleq\left[x_1',x_2',\ldots,x_n'\right]'\in\mathbb{C}^{mn}.
$$
Notice that $\mathrm{vec}(AXB)=(B'\otimes A)\mathrm{vec}(X)$.
For Kronecker product, we have $\left(A_1 A_2\right)\otimes\left(B_1 B_2\right)=
 \left(A_1\otimes B_1\right)\left(A_2\otimes B_2\right)$.
Take expectations and vectorization operators over both sides of~\eqref{eqn:peak-Pk-3}.
We obtain from Lemma~\ref{lemma:stopping-time-lemma} that
\begin{eqnarray}\label{eqn:vec-Pk}
\mathbb{E}[\mathrm{vec}(P_{\beta_{j+1}})]
&=&\mathbb{E}[\mathrm{vec}(W)]\\
&&\hspace{-15mm}+
\sum_{i=1}^\infty (A\otimes A)^i(1-q)^{i-1}q\sum_{l=1}^{\mathsf{I_o}-1}
\overline{(A^l+K^{(l)}C^{(l)})}\otimes (A^l+K^{(l)}C^{(l)})
p(1-p)^{l-1} \,\mathbb{E}[\mathrm{vec}(P_{\beta_j})].\notag
\end{eqnarray}
In the above equation $\mathbb{E}[\mathrm{vec}(W)]$ can be written as
\begin{eqnarray}\label{eqn:peak-Pk-constant}
\mathbb{E}[\mathrm{vec}(W)]
&=&\sum_{i=1}^\infty (A\otimes A)^i(1-q)^{i-1}q\,
\mathrm{vec}(U)+\sum_{i=1}^\infty \sum_{l=0}^{i-1}(A\otimes A)^l(1-q)^{i-1}q\,\mathrm{vec}(Q).
\end{eqnarray}
In Lemma~\ref{lemma:matrix-series-convergence}, we show that both of the two terms in~\eqref{eqn:peak-Pk-constant} converges if
$|\lambda_A|^2(1-q)<1$.

For $j=1$, following the similar argument as above, we have
$$\mathbb{E}\|P_{\beta_1}\|\leq \mathbb{E}\|P_{\tau_1}\|
\sum_{i=1}^\infty \|A^i\|^2 (1-q)^{i-1}q+
\Big\|\sum_{i=1}^\infty (1-q)^{i-1}q V_i\Big\|,
$$
where $V_i$'s are defined in~\eqref{eqn:peak-Pk-1}.
Moreover, by Lemma~\ref{lemma:riccati>Io} and~\eqref{eqn:h-geq-g}, it holds that
\begin{eqnarray*}
\mathbb{E}\|P_{\tau_1}\|&\leq&
\|\mathsf{L}I\|+\sum_{i=1}^{\mathsf{I_o}}\|g^i(\Sigma_0)\|(1-p)^{i-1}p\\
&\leq&\|\mathsf{L}I\|+ \|\Sigma_0\|\sum_{i=1}^{\mathsf{I_o}}\|A^i\|^2(1-p)^{i-1}p
+\|V_{\mathsf{I_o}}\|,
\end{eqnarray*}
showing that $\mathbb{E}\|P_{\tau_1}\|$ is bounded. To sum up,
$\mathbb{E}\|P_{\beta_1}\|$ is bounded if $|\lambda_A|^2(1-q)<1$.
By applying the Cauchy-Schwarz inequality to the inner product
of random variables, the boundness of $\mathbb{E}\|P_{\beta_1}\|$
implies the boundness of each element of $\mathbb{E}[P_{\beta_1}]$. So is
$\mathbb{E}[\mathrm{vec}(P_{\beta_1})]$ if $|\lambda_A|^2(1-q)<1$.


We have shown that
$\mathbb{E}[\mathrm{vec}(P_{\beta_{j}})]$ for $j\in\mathbb{N}$ evolves following~\eqref{eqn:vec-Pk}, and that $\mathbb{E}[\mathrm{vec}(W)]$
in~\eqref{eqn:vec-Pk} and $\mathbb{E}[\mathrm{vec}(P_{\beta_1})]$ are bounded if
$|\lambda_A|^2(1-q)<1$.
We conclude that if $|\lambda_A|^2(1-q)<1$ and there exists an
$K\triangleq [K^{(1)},\ldots,K^{(\mathsf{I_o}-1)}]$ such that $|\lambda_{H(K)}|$
defined in~\eqref{eqn:main-sufficient-cond} is less than $1$, then
the spectral radius of
$$\sum_{i=1}^\infty (A\otimes A)^i(1-q)^{i-1}q\sum_{l=1}^{\mathsf{I_o}-1}
\overline{(A^l+K^{(l)}C^{(l)})}\otimes (A^l+K^{(l)}C^{(l)}) (1-p)^{l-1}p$$
 is less than $1$, all the above observations lead to $\sup_{j\geq 1}\mathbb{E}[\mathrm{vec}_i(P_{\beta_j})]<\infty~\forall 1\leq i\leq n^2$,
where $\mathrm{vec}_i(X)$ represents the $i$th element of $\mathrm{vec}(X)$.
In addition, there holds  $$\mathbb{E}\|P_{\beta_j}\|\leq
\mathbb{E}\left[\mathrm{tr}(P_{\beta_j})\right]=
[e_1',\ldots,e_n']\,\mathbb{E}[\mathrm{vec}(P_{\beta_j})],$$ where
$e_i$ denotes the vector with a $1$ in the $i$th coordinate and 0's elsewhere,
so the desired result follows.

\section*{Appendix C. Proof of Theorem~\ref{thm:Contraction2LMI}}
\noindent $(i)\Rightarrow (ii)$. It suffices to show $|\lambda_{H^*(K)}|<1$.
The hypothesis means that for any $X\in \mathbb{S}_+^n$
\begin{equation}\label{eqn:limit-zero}
\lim_{k\rightarrow \infty} \left(H^*(K)\right)^k\mathrm{vec}(X)=0.
\end{equation}
In light of Lemma~\ref{lemma:matrix-depomositation}, for any $G\in \mathbb{C}^{n\times n}$
there exist $G_1,G_2,G_3, G_4\in \mathbb{S}_+^n$ such that $G=(G_1-G_2)+(G_3-G_4)\mathbf{i}$.
It can been seen from~\eqref{eqn:limit-zero} that
\begin{eqnarray*}
&&\lim_{k\rightarrow \infty} \left(H^*(K)\right)^k\mathrm{vec}(G)\\
&=&
\lim_{k\rightarrow \infty} \left(H^*(K)\right)^k\Big(\mathrm{vec}(G_1)-\mathrm{vec}(G_2)
+\mathrm{vec}(G_3)\mathbf{i}-\mathrm{vec}(G_4)\mathbf{i}\Big)\\
&=&0,
\end{eqnarray*}
which implies $(ii)$.

\noindent  $(ii)\Rightarrow (iii)$.
Since $|\lambda_{H^*(K)}|<1$ by the hypothesis in $(ii)$,
$(I-H^*(K))^{-1}$ exist and it equals to $\sum_{i=0}^{\infty}\left(H^*(K)\right)^i$.
Due to the nonsingular of $(I-H^*(K))^{-1}$ and the
one-to-one correspondence of vectorization operator,
for any positive definite matrix $V\in\mathbb{C}^{n\times n}$, there exists a unique matrix $P\in\mathbb{C}^{n\times n}$ such that
\begin{equation}\label{eqn:v-p-Lk}
\mathrm{vec}(V)=\left(I-H^*(K)\right)\mathrm{vec}(P).
\end{equation}
The property of Kronecker product gives
$\mathrm{vec}(V)=\mathrm{vec}\left(P-\mathcal{L}_K(P)\right).$
Since vectorization is one-to-one correspondence, we then have
$V=P-\mathcal{L}_K(P)>0.$
It still remains to show $P>0$.
It follows from~\eqref{eqn:v-p-Lk} that
\begin{eqnarray}\label{eqn:v-p-Lk}
\mathrm{vec}(P)&=&\left(I-H^*(K)\right)^{-1}\mathrm{vec}(V)\notag\\
&=&\sum_{i=0}^{\infty}\left(H^*(K)\right)^i\mathrm{vec}(V)\notag\\
&=&\mathrm{vec}\left(\sum_{i=0}^\infty \mathcal{L}_K^i(V)\right),
\end{eqnarray}
which yields
$
P=\sum_{i=0}^\infty \mathcal{L}_K^i(V)>0.
$

\noindent  $(iii)\Rightarrow (i)$.
If there exist $\,K=[K^{(1)},\ldots,K^{(\mathsf{I_o}-1)}]$ with each matrix
$K^{(i)}$ having compatible dimensions and  $P>0$ such that $\mathcal{L}_K(P)<P$, then there must exist a $\mu \in (0,1)$ satisfying
$\mathcal{L}_K(P)<\mu P$.
Choose an $c>0$ such that
$X\leq c P$.
Then, due to the linearity and non-decreasing properties
of $\mathcal{L}_K(X)$ with respect to $X$ on the positive
semi-definite cone,
for $k\in\mathbb{N}$
$$
\mathcal{L}^k_K(X)\leq \mathcal{L}^k_K(cP)=
c\mathcal{L}^k_K(P)
<c\mathcal{L}^{k-1}_K(\mu P)
<\cdots< c \mu^k P,
$$
which leads to $\lim_{k \rightarrow \infty}\mathcal{L}^k_K(X)=0$.

\noindent  $(iii)\Rightarrow (iv)$. It can be seen from $(iii)$ that
\begin{equation}\label{eqn:linear-operator}
p\sum_{i=1}^{\mathsf{I_o}-1}(1-p)^{i-1}(A^i+K^{(i)}C^{(i)})^*
Y(A^i+K^{(i)}C^{(i)})<P,~~P>0
\end{equation}
and $Y$ is the solution to the following
Lyapunov equation
\begin{equation}\label{eqn:Lypunov-eqn}
Y=(1-q)A'YA+qA'PA,
\end{equation}
where $qA'PA>0$ due to $q>0$, $P>0$ and non-singularity of $A$.
In light of the Schur Complement lemma,~\eqref{eqn:linear-operator}
is equivalent to
\begin{equation*}
\Xi\triangleq
\left[\begin{array}{ccccccc}
P& \sqrt{p}(A+KC)^* & \cdots &
\sqrt{p(1-p)^{\mathsf{I_o}-2}}(A^{\mathsf{I_o}-1}+K^{(i)}C^{(i)})^*\\
* & Y^{-1} &\cdots & 0\\
\vdots & \vdots &\ddots & \vdots\\
* &
* &\cdots & Y^{-1}
\end{array}\right]>0.
\end{equation*}
Then we obtain
\begin{eqnarray*}
&\left[\begin{array}{cccc}
I& 0 & \cdots &0\\
0 & Y &\cdots & 0\\
\vdots & \vdots &\ddots & \vdots\\
0 &0 &\cdots & Y
\end{array}\right]\Xi\left[\begin{array}{cccc}
I& 0 & \cdots &0\\
0 & Y &\cdots & 0\\
\vdots & \vdots &\ddots & \vdots\\
0 &0 &\cdots & Y
\end{array}\right]=\Psi.
\end{eqnarray*}
The equality holds by letting $F_i=(YK^{(i)})^*,~i=1,\dots,\mathsf{I_o}-1$, thereby~\eqref{eqn:LMI-1} follows.
By relaxing the equality in~\eqref{eqn:Lypunov-eqn} into inequality and applying the same method as above,~\eqref{eqn:LMI-2} follows.

\noindent  $(iv)\Rightarrow (iii)$. Note that, by
the Schur complement lemma and $X,Y>0$,~\eqref{eqn:LMI-2} holds if and only if
\begin{equation}\label{eqn:lypunov-ineq}
Y\geq (1-q)A'YA+qA'XA.
\end{equation}
Similarly,~\eqref{eqn:LMI-1} holds if and only if
\begin{equation}\label{eqn:linear-operator-ineq}
p\sum_{i=1}^{\mathsf{I_o}-1}(1-p)^{i-1}(A^i+K^{(i)}C^{(i)})^*Y(A^i+K^{(i)}C^{(i)})<X,
\end{equation}
where $K^{(i)}=Y^{-1}F_i^*,~i=1,\ldots,\mathsf{I_o}-1$.
Applying the inequality of~\eqref{eqn:lypunov-ineq} for $k$ times, it results in
$$Y\geq (1-q)^k(A')^kYA^k+q\sum_{j=1}^k(1-q)^{j-1}(A')^jXA^j.$$
As $Y$ is bounded, taking limitation on the right sides, it yields
\begin{equation}\label{eqn:Y-lyapunov-solution}
Y\geq q\sum_{j=1}^\infty(1-q)^{j-1}(A')^jXA^j.
\end{equation}
Combining~\eqref{eqn:linear-operator-ineq} and~\eqref{eqn:Y-lyapunov-solution}, we obtain
$\mathcal{L}_K(X)<X$, where $(iii)$ follows.

\section*{Appendix D. Proof of Theorem~\ref{thm:peak-2-usual-stability}}
To prove this theorem, we need some preliminary lemmas.
\begin{lemma}\label{lemma:unif-bound-pre-lemma}
Suppose that there exist constants $d_1,d_0\geq 0$ such that, for any $j\in\mathbb{N}$ and
$k\in [\beta_j,\beta_{j+1}]$,
$\|P_k\|\leq \max\limits_{i=j,j+1}\{d_1\|P_{\beta_{i}}\|+d_0\}$ holds $\mu-$almost surely. If $\sup_{j\in\mathbb{N}}\mathbb{E}\|P_{\beta_j}\|<\infty$, then
$\sup_{k\in\mathbb{N}}\mathbb{E}\|P_{k}\|<\infty$ holds.
\end{lemma}
{\it Proof.}
Since $\sup_{j\in\mathbb{N}}\mathbb{E}\|P_{\beta_j}\|<\infty$, there
exists a uniform bound $\alpha$ for $\{\mathbb{E}\|P_{\beta_j}\|\}_{j\in \mathbb{N}}$,
i.e., $\mathbb{E}\|P_{\beta_j}\|\leq \alpha$ for all $j\in \mathbb{N}$. By the definition of
$\beta_j$ in~\eqref{def:beta-j},
$k$ should be no larger than $\beta_k$ for all $k\in \mathbb{N}$.
Then one obtains
\begin{eqnarray*}
\mathbb{E}\|P_k\|&=&\mathbb{E}\left[\sum_{j=0}^{k-1}
\mathbb{E}\Big[\|P_k\|\mid\beta_j\leq k\leq \beta_{j+1}\Big]1_{\{\beta_j\leq k\leq \beta_{j+1}\}}\right]\\
&\leq&\sum_{j=0}^{k-1}
\mathbb{E}\left[\max_{i=j,j+1}\{d_1\|P_{\beta_{i}}\|+d_0\}
\mid\beta_j\leq k\leq \beta_{j+1}\right]
\mu(\beta_j\leq k\leq \beta_{j+1})\\
&\leq&\sum_{j=0}^{k-1}
\left(d_1\mathbb{E}\|P_{\beta_j}\|+d_1\mathbb{E}\|P_{\beta_{j+1}}\|+d_0\right)
\mu(\beta_j\leq k\leq \beta_{j+1})\\
&\leq&\left(2d_1\sup_{j\leq k}\mathbb{E}\|P_{\beta_j}\|+d_0\right)\sum_{j=0}^{k-1}\mu(\beta_j\leq k\leq \beta_{j+1})\\
&\leq&2d_1\alpha+d_0,
\end{eqnarray*}
which completes the proof.
\hfill$\square$

Before proceeding to the proof of the theorem,
let us provide some properties related to the
discrete-time algebraic Riccati equation (DARE).
The proof, provided in~\cite{lancaster1995ARE}, is omitted.
\begin{lemma}\label{lemma:DARE-stable}
Consider the following DARE
\begin{equation}\label{eqn:DARE}
P=APA'+Q-APC'(CPC'+R)^{-1}CPA'.
\end{equation}
If $(A,{Q}^{{1}/{2}})$ is controllable
and $(C,A)$ is observable, then it has a unique
positive definite solution $\tilde P$ and
$A+\tilde{K}C$ is stable, where $\tilde{K}=-A\tilde PC'(C\tilde PC'+R)^{-1}$.
\end{lemma}

Fix  $j\geq 0$. First of all, we shall show that, for
$k\in [\beta_j+1,\tau_{j+1}]$, $\|P_k\|$ is uniformly bounded by
an affine function of $\|P_{\beta_j}\|$.
By Lemma~\ref{lemma:gx} and Lemma \ref{lemma:DARE-stable},
we have $g(P_{k-1})\leq \phi_1(\tilde{K},P_{k-1})$ and that $A+\tilde{K}C$ is stable.
In light of \eqref{eqn:g-monotone} in Lemma~\ref{lemma:appendix-g-h}, we further obtain
$g^i(P_{k-1})\leq \phi_1^i(\tilde{K},P_{k-1})$ for all $i\in \mathbb{N}$.
Therefore, an upper bound of $\|P_k\|$ is given as follows:
\begin{eqnarray*}
\|P_k\|&=&\|g^{k-\beta_j}(P_{\beta_j})\|\\
&\leq &\|\phi_1^{k-\beta_j}(\tilde{K},P_{\beta_j})\|\\
&\leq&\Big\|(A+\tilde{K}C)^{k-\beta_j}P_{\beta_j}(A'+C'\tilde{K}^*)^{k-\beta_j}\Big\|
+\Big\|\sum_{i=0}^{k-\beta_j-1}(A+\tilde{K}C)^i(Q+\tilde{K}R\tilde{K}^*)
(A'+C'\tilde{K}^*)^i\Big\|\\
&\leq&\|(A+\tilde{K}C)^{k-\beta_j}\|^2\|P_{\beta_j}\|+
\sum_{i=0}^{k-\beta_j-1}\|(A+\tilde{K}C)^i\|^2\|Q+\tilde{K}R\tilde{K}^*\|\\
&\leq&m_0\,\alpha_0^{2k-2\beta_j}\|P_{\beta_j}\|+
m_0\|Q+\tilde{K}R\tilde{K}^*\|\sum_{i=0}^{k-\beta_j-1}\alpha_0^{2i},
\end{eqnarray*}
where
$\alpha_0=|\lambda_{A+\tilde{K}C}|+\epsilon_0\|A+\tilde{K}C\|$
 and
$m_0=n(1+2/\epsilon_0)^{2n-2}$ with
 a positive number $\epsilon_0$ satisfying
$|\lambda_{A+\tilde{K}C}|+\epsilon_0\|A+\tilde{K}C\|<1$ (such
an $\epsilon_0$ must exist because $|\lambda_{A+\tilde{K}C}|<1$), and the last inequality holds due to Lemma~\ref{lemma:norm2radius}.
Observe that $\sum_{i=0}^{k-\beta_j-1}\alpha_0^{2i}\leq
\frac{1}{1-\alpha_0^2}$.
As $\alpha_0<1$, $\alpha_0^{2k-2\beta_j}<1$ for any $k\in[\beta_j+1,\tau_{j+1}]$.
Therefore,
\begin{equation}\label{eqn:unifrom-bound1}
\|P_k\|\leq m_0\|P_{\beta_j}\|+n_0,
\end{equation}
where $n_0\triangleq\frac{m_0}{1-\alpha_0^2}\|Q+\tilde{K}R\tilde{K}'\|$.

Next, we shall show that, for $k\in [\tau_{j+1}+1,\beta_{j+1}]$, $\|P_k\|$ is
bounded by an affine function of $\|P_{\beta_{j+1}}\|$. To do this,
let us look at the relationship between $P_{\beta_{j+1}}$ and $P_k$. Since
$\gamma_k=0$ for all $k\in [\tau_{j+1},\beta_{j+1}-1]$, the relation is
given by
$$P_{\beta_{j+1}}=A^{\beta_{j+1}-j}P_k (A')^{\beta_{j+1}-k}
+\sum_{i=0}^{\beta_{j+1}-k-1}A^iQ(A')^i,$$
from which we obtain
$P_{\beta_{j+1}}\geq A^{\beta_{j+1}-k}P_k (A')^{\beta_{j+1}-k}.$
Then it yields
\begin{eqnarray*}
\|P_{\beta_{j+1}}\|&\geq &\|A^{\beta_{j+1}-k}P_k (A')^{\beta_{j+1}-k}\|\\
&\geq& \frac{1}{n}\mathrm{Tr}(A^{\beta_{j+1}-k}P_k (A')^{\beta_{j+1}-k})\\
&=&\frac{1}{n}\mathrm{Tr}(P_k^{{1}/{2}} (A')^{\beta_{j+1}-k}A^{\beta_{j+1}-k}P_k^{{1}/{2}})\\
&\geq&\frac{1}{n}\|A^{k-\beta_{j+1}}\|^{-2}\mathrm{Tr}(P_j)\\
&\geq&\frac{1}{n}\|A^{k-\beta_{j+1}}\|^{-2}\|P_k\|,
\end{eqnarray*}
where the second and the last inequality allows from the fact that $\|X\|=\lambda_{X}\geq \frac{1}{n}\mathrm{Tr}(X)$ and $\mathrm{Tr}(X)\geq \|{X}\|$ for any $X\in\mathbb{S}_+^n$;
the third one holds since $$(A')^{\beta_{j+1}-k}A^{\beta_{j+1}-k}\geq
\min \sigma\big(A^{\beta_{j+1}-k}(\cdot)'\big)I=\frac{1}{\lambda_{A^{k-\beta_{j+1}}(\cdot)'}}I
=\|A^{k-\beta_{j+1}}\|^{-2}I.$$
If $A$ has no eigenvalues on the unit circle, by Lemma~\ref{lemma:norm2radius}, there holds $\|A^{k-\beta_{j+1}}\|\leq n_1\, \alpha_1^{\beta_{j+1}-k}$ where
$n_1\triangleq\sqrt{n}(1+2/\epsilon_1)^{n-1}$ and $\alpha_1\triangleq
|\lambda_{A^{-1}}|+\epsilon_1\|A^{-1}\|$ with a positive number $\epsilon_1$ so that $\alpha_1<1$ (such an $\epsilon_1$ must exist
since $|\lambda_{A^{-1}}|< 1$ by assumption~\ref{asmpt:assumpt-A}).
As $\alpha_1<1$, $\|A^{k-\beta_{j+1}}\|\leq n_1\, \alpha_1^{\beta_{j+1}-k}<n_1$ for all $k\in[\beta_j+1,\tau_{j+1}]$.
If $A$ has semi-simple eigenvalues on the unit circles, we denote the Jordan form
of $A$ as $J=\mathrm{diag}(J_{11},J_{22})$, where
$A_{11}$ has no eigenvalues on the unit circle and $A_{22}$ is diagonal with
all semi-simple eigenvalues on the unit circle, i.e., there exists a nonsingular matrix $S\in\mathbb{R}^{n\times n}$ such that
$J=SAS^{-1}.$
In this case,
$$
\|J^{k-\beta_{j+1}}\|=\max\left\{\|J_{11}^{k-\beta_{j+1}}\|,\|J_{22}^
{k-\beta_{j+1}}\|\right\}
\leq \max\{n_1,1\}\leq n_1,
$$
where, similarly, $n_1\triangleq\sqrt{n}(1+2/\epsilon_1)^{n-1}$ with
a positive number $\epsilon_1$ so that $|\lambda_{J_{11}^{-1}}|+\epsilon_1\|J_{11}^{-1}\|<1$.
Since $\|S^{-1}\cdot S\|$ can be considered a matrix norm, we have
$$
\|A^{k-\beta_{j+1}}\|=\|S^{-1}J^{k-\beta_{j+1}}S\|\leq
c\|J^{k-\beta_{j+1}}\|\leq cn_1,
$$
where $c=\sup_{X\in\mathbb{C}^{n\times n}}\frac{\|S^{-1}XS\|}{\|X\|}<\infty$
due to the equivalence of matrix norms on a finite dimensional vector space.
Then, we have the following upper bound for $\|P_k\|$ for all $k\in[\tau_{j+1}+1,\beta_{j+1}]$:
\begin{equation}\label{eqn:unifrom-bound2}
\|P_k\|\leq n\|A^{k-\beta_{j+1}}\|^2\|P_{\beta_{j+1}}\|\leq m_1\|P_{\beta_{j+1}}\|,
\end{equation}
where $m_1\triangleq nc^2n_1^2$.

According to~\eqref{eqn:unifrom-bound1} and~\eqref{eqn:unifrom-bound2},
it can be seen that when $k\in[\beta_j,\beta_{j+1}]$ $$\|P_k\|\leq\max\{m_0,m_1\}\max\{\|P_{\beta_{j}}\|,\|P_{\beta_{j+1}}\|\}+n_0.$$
Since $j$ is arbitrarily chosen, by invoking Lemma~\ref{lemma:unif-bound-pre-lemma}, the desired conclusion follows.

\section*{Appendix E. Proof of Proposition~\ref{thm:critical-value-p}}

If $p=0$, we have the standard Kalman filter, which evidently converges to a bounded estimation error covariance. On the other hand, if $p=1$, then the Kalman filter reduces to an open-loop predictor after time step $k=2$, which
 suggests that there exists a transition point for $p$ beyond which
 the expected prediction error covariance matrices are not uniformly bounded.
It remains to show that with a given $q$ this transition point is unique. Fix a $0\leq p_1<1$ such that $\sup_{k\in\mathbb{N}}\mathbb{E}_{p_1}\|P_k\|<\infty~\forall \Sigma_0\geq 0$. It suffices to show that, for any $p_2<p_1$, $\sup_{k\in\mathbb{N}}\mathbb{E}_{p_2}\|P_k\|<\infty$ for all $\Sigma_0\geq 0$.
To differentiate two Markov chains with different failure rate in~\eqref{eqn:markov-trnsition-prob}, we use the notation $\{\gamma_k(p_i)\}_{k\in\mathbb{N}}$ instead to represent the packet loss process so as to
indicate the configuration $p=p_i$ in~\eqref{eqn:markov-trnsition-prob}.
We will prove the aforementioned statement using a coupling argument.
We define a sequence of random vectors $\{(z_k,\tilde z_k)\}_{k\in\mathbb{N}}$
over a probability space $(\mathscr{G},\mathcal{G},\pi)$
with $\mathscr{G}=\{(0,0),(0,1),(1,1)\}^{\mathbb{N}}$ and $\mathcal{G}_k$ representing the filtration generated by $(z_1,\tilde z_1),\ldots,(z_k,\tilde z_k)$.
We also define
$$\varphi_{1}(\{z_k,\tilde z_k\}_{k=1:t})=\psi_{z_t}\circ\cdots\circ
\psi_{z_1}(\Sigma_0)
$$
and
$$\varphi_{2}(\{z_k,\tilde z_k\}_{k=1:t})=\psi_{\tilde z_t}\circ\cdots\circ
\psi_{\tilde z_1}(\Sigma_0),$$
where $\psi_z=z g+(1-z)h$ with $h,g$ defined in~\eqref{eqn:h-func},~\eqref{eqn:g-func} and $z=\{0,1\}$.
Due to Lemma~\ref{lemma:appendix-g-h} and $z_k\leq\tilde z_k$ in $\mathscr{G}$,
we have $\varphi_{1}\left(\{z_k,\tilde z_k\}_{k=1:t}\right)\geq \varphi_{2}\left(\{z_k,\tilde z_k\}_{k=1:t}\right)$.

\begin{figure} 
      \centering
      \includegraphics[width=3in]{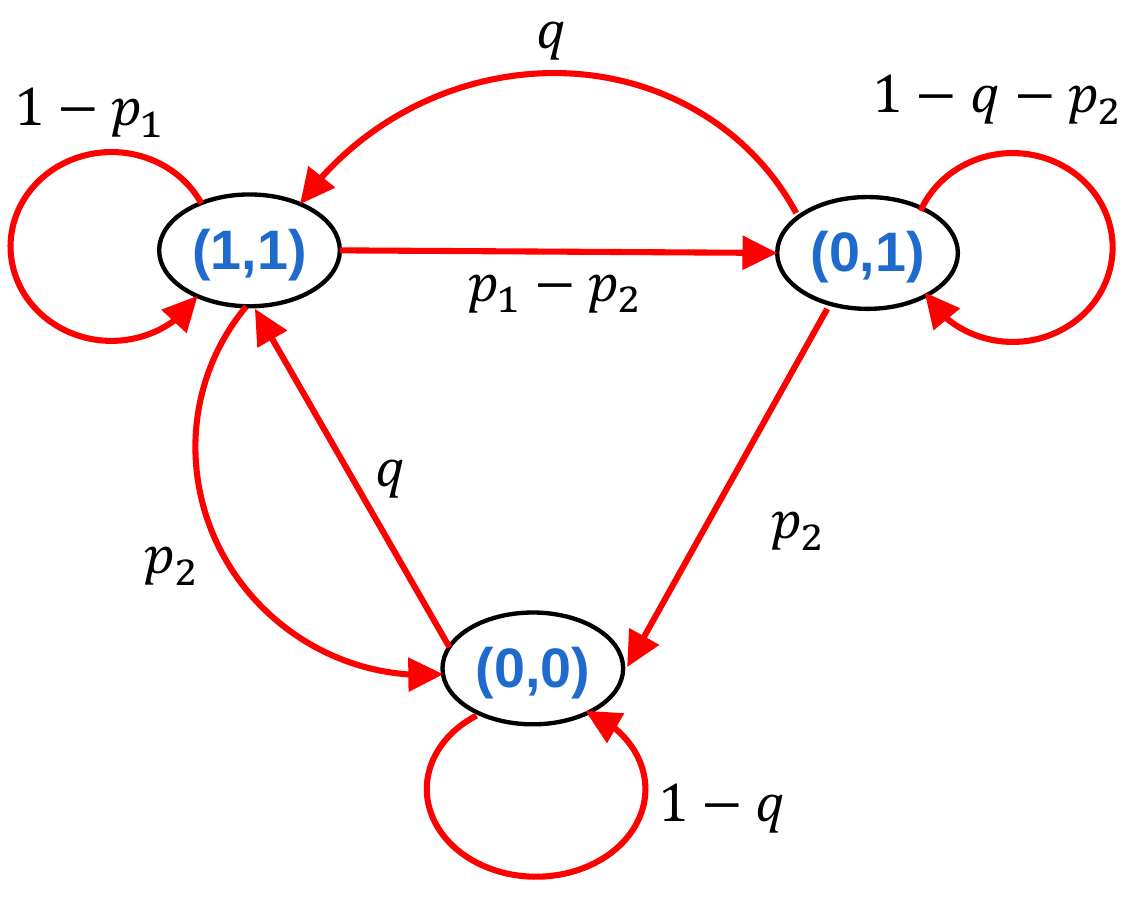}
  \caption{The transitions of the Markov chain $\{(z_k,\tilde z_k)\}_{k\in\mathbb{N}}$ when $p_2+q\leq 1$.}
  \label{Fig:Markov-chains-p}
\end{figure}

When $p_2+q\leq 1$, we let the evolution of $\{(z_k,\tilde z_k)\}_{k\in\mathbb{N}}$
follow the Markov chain illustrated in Fig.~\ref{Fig:Markov-chains-p}, whereby
it can seen that $\pi(z_{k+1}=j|z_k=i)$'s for $i,j=\{0,1\}$ are constants independent of $\tilde z_k$'s,
and conversely that $\pi(\tilde z_k=j|\tilde z_k=i)$'s for $i,j=\{0,1\}$ are constants independent of $z_k$'s. Therefore, both the marginal distributions of $\{z_k\}_{k\in\mathbb{N}}$ and $\{\tilde{z}_k\}_{k\in\mathbb{N}}$ are Markovian, and moreover, $$\pi(z_{k+1}=j|z_k=i)=
 \mathbb{P}_{p_1}(\gamma_{k+1}(p_1)=j|\gamma_k(p_1)=i)$$
 and
 $$\pi(\tilde z_{k+1}=j|\tilde z_k=i)=
 \mathbb{P}_{p_2}(\gamma_{k+1}(p_2)=j|\gamma_k(p_2)=i)$$
 for all $i,j=\{0,1\}$ and $k\in\mathbb{N}$.
  It can be seen that the Markov chain in Fig.~\ref{Fig:Markov-chains-p}
is ergodic and has a unique stationary distribution
\begin{equation}\label{eqn:mc-stationary-dist}
\pi_\infty\left((0,0)\right)=\frac{p_2}{p_2+q},~\;\;
\pi_\infty\left((0,1)\right)=\frac{p_1}{p_1+q}-\frac{p_2}{p_2+q},~\;\;
\pi_\infty\left((1,1)\right)=\frac{q}{p_1+q}.
\end{equation}
We assume the Markov chain starts at the stationary distribution.
Then the distribution of $(z_k,\tilde z_k)$ for $k\geq 2$ is the same as
$(z_1,\tilde z_1)$, which gives
 \begin{eqnarray*}
 \mathbb{E}_{p_1}^{\infty}\|P_k\|&=&
 \int_{\Omega}\left\|\psi_{\gamma_k(p_1)}\circ\cdots\circ
 \psi_{\gamma_1(p_1)}(\Sigma_0)\right\| \,\rm{d} \mathbb{P}_{p_1}\\
 &=&\int_{\mathscr{G}}\big\|\varphi_{1}(\{z_j,\tilde z_j\}_{j=1:k})\big\| \,\rm{d}{\pi}\\
 &\geq&\int_{\mathscr{G}}\big\|\varphi_{2}(\{z_j,\tilde z_j\}_{j=1:k})\big\| \,\rm{d}{\pi}\\
 &=&\int_{\Omega}\left\|\psi_{\gamma_k(p_2)}\circ\cdots\circ
 \psi_{\gamma_1(p_2)}(\Sigma_0)\right\| \,\rm{d} \mathbb{P}_{p_2}\\
 &=&\mathbb{E}_{p_2}^{\infty}\|P_k\|,
 \end{eqnarray*}
where $\mathbb{E}^\infty$ means that the expectations is taken conditioned
on the stationarily distributed $\gamma_1$.

When $p_2+q>1$, we abuse the definition of probability measure
and allow the existence of negative probabilities in the
Markov chain described in Fig.~\ref{Fig:Markov-chains-p}, generating  $\{(z_k,\tilde z_k)\}_{k\in\mathbb{N}}$.
It can be easily shown by direct computation  that the eigenvalues of transition probability matrix, denoted by $\mathbf{M}\in\mathbb{R}^{3\times 3}$,  of this Markov chain are $1-q-p_1,~1-q-p_2$ and $1$, respectively.
As a result, $\mathbf{M}^k$ converges to a limit as $k$ tends to infinity,   indicating that
the generalized Markov chain has  a unique stationary distribution which is the  same as
the one given in~\eqref{eqn:mc-stationary-dist}.
Therefore, although the Markov chain is only formally defined  without corresponding physical meaning with $p_2+q>1$,
the most important basic property for  this coupling still persists, that is,
$$\pi(z_1=i_1,\ldots,z_t=i_t)=\mathbb{P}_{p_1}(\gamma_1=i_1,\ldots,
\gamma_t=i_t)$$
and
$$\pi(\tilde z_1=i_1,\ldots,\tilde z_t=i_t)=\mathbb{P}_{p_2}(\gamma_1=i_1,\ldots,\gamma_t=i_t)$$
for all $t\in\mathbb{N}$ and $i_1,\ldots,i_t\in\{0,1\}$.
Thus, the inequality $\mathbb{E}_{p_1}^{\infty}\|P_k\|\geq  \mathbb{E}_{p_2}^{\infty}\|P_k\|$ still proves true in this case.


Finally, in order to show $\sup_{k\in\mathbb{N}}\mathbb{E}_{p_2}\|P_k\|<\infty$ with
packet losses initialized by $\gamma_1=1$, we only need to recall the following lemma:
\begin{lemma}[Lemma 2 in~\cite{you2011mean}]
The statement holds that  $\sup_{k\in\mathbb{N}}\mathbb{E}^{\infty}\|P_k\|<\infty$
if and only if $\sup_{k\in\mathbb{N}}\mathbb{E}^{1}\|P_k\|<\infty$
and $\sup_{k\in\mathbb{N}}\mathbb{E}^{0}\|P_k\|<\infty$, where
$\mathbb{E}^1$ and $\mathbb{E}^0$ denotes the expectations conditioned
on $\gamma_1=1$ and $0$,  respectively.
\end{lemma}
The proof is now complete.

\section*{Appendix F. Proof of Theorem~\ref{proposition:critical-curve}}

Let $p=p_c(q)$ be the critical value established in Proposition~\ref{thm:critical-value-p}.
For any given $p$, fix
a $1-1/|\lambda_A|^2<q_1<1$ so that $\sup_{k\in\mathbb{N}}\mathbb{E}_{q_1}\|P_k\|<\infty$
for all $\Sigma_0\geq 0$. From a symmetrical coupling argument  as the proof of Proposition~\ref{thm:critical-value-p}, for any $q_1\leq q_2<1$, $\sup_{k\in\mathbb{N}}\mathbb{E}_{q_2}\|P_k\|<\infty$ also holds
for all $\Sigma_0\geq 0$. As a result, $p_c(q)$ is a non-decreasing  function of $q$.

Consequently, $p_c(\cdot)$ yields an inverse function, denoted $q_c(\cdot)$, which is also non-decreasing. The desired conclusion then follows immediately, e.g.,  we can simply choose $f_c(p,q)=p_c(q)-p$.

\section*{Appendix G. Proof of Theorem~\ref{thm:sufficient-cond}}
We shall first show that
\begin{equation}\label{eqn:eta-LMI}
\lim_{\eta\rightarrow 1^+}{p}(\eta)=\underline{p}
\end{equation}
holds where $\eta>1$ is properly taken so that $\eta^2|\lambda_A|^2(1-q)<1$,
and
$${p}(\eta)
\triangleq \sup \big\{p:
\exists (K,P)\hbox{~s.t.~}\mathcal{L}(\eta A, K,P,p)<P,P>0\big\}$$ with the notation  $\mathcal{L}(A, K,P,p)$ used to alter $\mathcal{L}_K(P)$ in the proof so as to emphasize
the relevance of $\mathcal{L}_K(P)$ to $A$ and $p$.
To this end, first note that ${p}(\eta)$
is a non-increasing function of $\eta$. Thus,
$\lim_{\eta\rightarrow 1^+}{p}(\eta)$
must exist.
To show the equality in~\eqref{eqn:eta-LMI}, we require the following lemma.
\begin{lemma}\label{lemma:contiuity-Lyapunov}
Suppose $X,\tilde X>0$ are the solutionis to the following Lyapunov equations respectively:
$$X=(1-q)AXA'+\tilde Q,~~~\eta^{-1}\tilde X=(1-q)A\tilde XA'+\tilde Q,$$
where $\tilde Q>0$, $0<q<1$ and $\eta>1$ are properly taken so that $\eta(1-q)|\lambda_A|^2<1$. Then, for any $\epsilon>0$ there always exists a $\delta>0$ such that $\eta\leq 1+\delta$ implies $\tilde X\leq (1+\epsilon)X.$
\end{lemma}
{\it Proof.} First we shall find an upper bound for $X$ and a lower bound for
$\tilde X$. It is straightforward that
\begin{equation}\label{eqn:X-lower-bound}
X\geq \tilde Q\geq  \|{\tilde Q}^{-1}\|^{-1}I.
\end{equation}
Let $d\triangleq \left[{(1-q)|\lambda_A|^2}\right]^{-{1}/{4}}>1$ and
restrict $1<\eta\leq d$.
By Lemma~\ref{lemma:norm2radius}, we have
\begin{eqnarray*}
\tilde X &=&\sum_{i=0}^{\infty} \eta^i(1-q)^iA^i\tilde Q(A^i)'\\
&\leq& \|\tilde Q\|\sum_{i=0}^{\infty} \eta^i(1-q)^i\|A^i\|^2I\\
&\leq& n \|\tilde Q\| (1+2/\varepsilon)^{2n-2}
\sum_{i=0}^{\infty} d^i(1-q)^i(|\lambda_A|+
\varepsilon \|A\|)^{2i}I
\end{eqnarray*}
for any $\varepsilon>0.$
Taking $\varepsilon=\frac{(d-1)|\lambda_A|}{\|A\|}$, it yields that
$\tilde X\leq\frac{cd }{d-1}\|\tilde Q\|$ with
$c=n (1+2/\varepsilon)^{2n-2}$.
Note that $\tilde X-X$ is bounded in the following way
\begin{eqnarray*}
\tilde X-X&=&\eta(1-q)A\tilde X A' +\eta  \tilde Q
-(1-q)AX A' - \tilde Q\\
&=&(1-q)A(\tilde X-X) A' +
(\eta-1)\left[
(1-q)A\tilde X A'+ \tilde Q\right]\\
&\leq&(1-q)A(\tilde X-X) A'+
(\eta-1)\tilde X\\
&\vdots&\\
&\leq&(1-q)^{k}A^{k}(\tilde X-X)(\cdot)'+
(\eta-1)\sum_{i=0}^{k-1}(1-q)^iA^i\tilde X (A^i)'
\end{eqnarray*}
Taking limitation on both sides, we obtain that
$\tilde X-X\leq (\eta-1)\sum_{i=0}^\infty(1-q)^iA^i\tilde X (A^i)'$,
which by Lemma~\ref{lemma:norm2radius} and~\eqref{eqn:X-lower-bound} gives
\begin{eqnarray*}
\tilde X-X&\leq&
(\eta-1)\|\tilde Q\|\frac{cd }{d-1}\sum_{i=0}^\infty(1-q)^i\|A^i\|^2\,I\\
&\leq& \frac{c^2 d^3 }{(d^2-1)(d-1)}(\eta-1)\|\tilde Q\|\,I\\
&\leq&\frac{c^2 d^3}{(d^2-1)(d-1)}(\eta-1)
\|\tilde Q\|\|{\tilde Q}^{-1}\|\,X\\
\end{eqnarray*}
where the last inequality holds because of~\eqref{eqn:X-lower-bound}.
Due to the positive definiteness of $\tilde Q$,
the assertion follows by letting $$1<\eta\leq \min\left\{\frac{(d^2-1)(d-1)
}{c^2 d^3\|\tilde Q\|\|{\tilde Q}^{-1}\|}\epsilon+1,d\right\}.$$
\hfill$\square$

By the definition of $\underline{p}$ one can verify that for any $0<\varepsilon<\underline{p}$
there always exists at least a $p\in (\underline{p}-\varepsilon,\underline{p})$~so that there exist $K$ and $P>0$
satisfying $\mathcal{L}(A, K,P,p)<P$; otherwise it contradicts~\eqref{eqn:upper-bound-p}.
We take an $\epsilon>0$ so that
$(1+\epsilon)\mathcal{L}(A, K,P,p)<P$
still holds. Then, from Lemma~\ref{lemma:contiuity-Lyapunov},
there always exists an $\eta_0>1$ satisfying $\Phi_{\eta_0}\leq\sqrt{1+\epsilon}\Phi$
where $\Phi$ and $\Phi_{\eta_0}$ are the positive definite solutions to the following equations
respectively:
$$  \Phi=(1-q)A' \Phi A+
A'P A,~~~ \eta_0^{-2} \Phi_{\eta_0}=(1-q)A'\Phi_{\eta_0} A+
A'P A.
$$
In addition, there exists an $\eta_1>1$ such that $\eta_1^{2\mathsf{I_o}-2}\leq \sqrt{1+\epsilon}.$
Letting $\tilde{\eta}=\min\{\eta_0,\eta_1\}$, we have
$$
P>(1+\epsilon)\mathcal{L}(A, K,P,p)\geq
p\sum_{i=1}^{\mathsf{I_o}-1}(1-p)^{i-1}(\tilde{\eta}^iA^i+\tilde{\eta}^iK^{(i)}C^{(i)})^*
(\Phi_{\tilde{\eta}})(\tilde{\eta}^iA^i+\tilde{\eta}^iK^{(i)}C^{(i)}),
$$
which implies that $\mathcal{L}(\tilde{\eta}A, K_{\tilde{\eta}},P,p)<P$
with $K_{\tilde{\eta}}\triangleq[\tilde{\eta}K^{(1)},\ldots,\tilde{\eta}^{\mathsf{I_o}-1}K^{(\mathsf{I_o}-1)}]$
and therefor that
${p}(\tilde{\eta})>\underline{p}-\varepsilon$.
As $\varepsilon$ is any positive real number,
$\lim_{\eta\rightarrow 1^+}{p}(\eta)=\underline{p}$ consequently holds.
Since $\eta A$ has no defective eigenvalues on the unit circle,
combining Theorems~\ref{thm:Contraction2LMI}~and~\ref{thm:peak-2-usual-stability}, we obtain that $p_c(\eta A, C, q)\geq {p}(\eta)$ holds.

To conclude, we also need to establish  the following lemma.
\begin{lemma}\label{lemma:critical-value-critical-stable-eigen}
For a given recovery rate $q$ satisfying $|\lambda_A|^2(1-q)<1$, denote $q_c(A,C,q)$
as the critical value of $q_c$ for a system $(C,A)$. Then, we have
\begin{equation}\label{eqn:criticavalue-eigen-bound}
p_c(A,C,q)\geq \lim_{\eta\rightarrow 1^+}p_c(\eta A,C,q).
\end{equation}
\end{lemma}
{\it Proof.}
To emphasize the relevance of $h(X)$ and $g(X)$ to $A$,
we will change the notations $h(X)$ and $g(X)$ into $h(A,X)$ and $g(A,X)$
respectively in the proof.
Note that $h(\eta A,X)$ and $g(\eta A,X)$ are both non-decreasing
function of $\eta>1$, where $\eta$ is properly chosen so that
$\eta^2|\lambda_A|^2(1-q)<1$, since, for all $1\leq \eta_1\leq \eta_2\leq 1+\delta$ and $X\geq 0$, one has
$$
h(\eta_2A,X)-h(\eta_1A,X)=(\eta_2^2-\eta_1^2) AXA'\geq 0
$$
and
$$
g(\eta_2A,X)-g(\eta_1A,X)=(\eta_2^2-\eta_1^2)(AXA'-AXC'(CXC'+R)^{-1}CXA')\geq 0.
$$
According to the fact that $P_k=(1-\gamma_k)h(A,P_{k-1})+\gamma_k g(A,P_{k-1})$ and that
$h(A,X)$ and $g(A,X)$ are non-decreasing functions of $X$ from Lemma~\ref{lemma:appendix-g-h},
we can easily show by induction that $P_k$ is also non-decreasing with respect to
$\eta$. Therefore the limitation on the right side of~\eqref{eqn:criticavalue-eigen-bound}
always exists and then the conclusion follows.
\hfill$\square$

From Lemma~\ref{lemma:critical-value-critical-stable-eigen} and what has been proved previously, it can be seen that
$\lim_{\eta\rightarrow 1^+}p_c(\eta A,C,q)$ and $\lim_{\eta\rightarrow 1^+}{p}(\eta)$ exist and moreover that
\begin{equation*}
p_c(A,C,q)\geq \lim_{\eta\rightarrow 1^+}p_c(\eta A,C,q)\geq \lim_{\eta\rightarrow 1^+}{p}(\eta)=\underline{p},
\end{equation*}
whereby the desired result follows.

\bibliographystyle{IEEETran}

\bibliography{sj_reference}

\begin{thebibliography}{10}
\providecommand{\url}[1]{#1}
\csname url@samestyle\endcsname
\providecommand{\newblock}{\relax}
\providecommand{\bibinfo}[2]{#2}
\providecommand{\BIBentrySTDinterwordspacing}{\spaceskip=0pt\relax}
\providecommand{\BIBentryALTinterwordstretchfactor}{4}
\providecommand{\BIBentryALTinterwordspacing}{\spaceskip=\fontdimen2\font plus
\BIBentryALTinterwordstretchfactor\fontdimen3\font minus
  \fontdimen4\font\relax}
\providecommand{\BIBforeignlanguage}[2]{{%
\expandafter\ifx\csname l@#1\endcsname\relax
\typeout{** WARNING: IEEEtran.bst: No hyphenation pattern has been}%
\typeout{** loaded for the language `#1'. Using the pattern for}%
\typeout{** the default language instead.}%
\else
\language=\csname l@#1\endcsname
\fi
#2}}
\providecommand{\BIBdecl}{\relax}
\BIBdecl

\bibitem{Delchamps}
D.~F. Delchamps, ``Stabilizing a linear system with quantized state feedback,''
  \emph{IEEE Transactions on Automatic Control}, vol.~35, pp. 916--924, 1990.

\bibitem{wong1}
W.~S. Wong and R.~W. Brockett, ``Systems with finite communication
  bandwidth-part {I}: State estimation problems,'' \emph{IEEE Transactions on
  Automatic Control}, vol.~42, no.~9, pp. 1294--1299, 1997.

\bibitem{wong2}
------, ``Systems with finite communication bandwidth-part {II}: Stabilization
  with limited information feedback,'' \emph{IEEE Transactions on Automatic
  Control}, vol.~44, no.~5, pp. 1049--1053, 1999.

\bibitem{brockett1}
R.~W. Brockett and D.~Liberzon, ``Quantized feedback stabilization of linear
  systems,'' \emph{IEEE Transactions on Automatic Control}, vol.~45, no.~7, pp.
  1279--1289, 2000.

\bibitem{nair1}
G.~N. Nair and R.~J. Evans, ``Communication-limited stabilization of linear
  systems,'' in \emph{Proceedings of the 39th IEEE Conference on Decision and
  Control}, 2000, pp. 1005--1010.

\bibitem{tatikonda2}
S.~Tatikonda and S.~Mitter, ``Control under communication constraints,''
  \emph{IEEE Transactions on Automatic Control}, vol.~49, pp. 1056 -- 1068,
  July 2004.

\bibitem{elia}
N.~Elia and S.~Mitter, ``Stabilization of linear systems with limited
  information,'' \emph{IEEE Transactions on Automatic Control}, vol.~46, no.~9,
  pp. 1384--1400, 2001.

\bibitem{Elia_N}
N.~Elia, ``Remote stabilization over fading channels,'' \emph{System and
  Control Letters}, vol. 54(3), pp. 237--249, March 2005.

\bibitem{ishii}
H.~Ishii and B.~A. Francis, ``Quadratic stabilization of sampled-data systems
  with quantization,'' \emph{Automatica}, vol.~39, pp. 1793--1800, 2003.

\bibitem{fu-xie-tac05}
M.~Fu and L.~Xie, ``The sector bound approach to quantized feedback control,''
  \emph{IEEE Transactions on Automatic Control}, vol.~50, no.~11, pp. 1698 --
  1711, 2005.

\bibitem{joao07}
J.~Hespanha, P.~Naghshtabrizi, and Y.~Xu, ``A survey of recent results in
  networked control systems,'' \emph{Proceedings of the IEEE}, vol.~95, no.~1,
  pp. 138--162, 2007.

\bibitem{Sinopoli2004}
B.~Sinopoli, L.~Schenato, M.~Franceschetti, K.~Poola, M.~Jordan, and S.~Sastry,
  ``Kalman filtering with intermittent observations,'' \emph{IEEE Transactions
  on Automatic Control}, vol.~49, no.~9, pp. 1453--1464, 2004.

\bibitem{Plarre09tac}
K.~Plarre and F.~Bullo, ``On {K}alman filtering for detectable systems with
  intermittent observations,'' \emph{IEEE Transactions on Automatic Control},
  vol.~54, no.~2, pp. 386--390, Feb 2009.

\bibitem{yilin10bound}
Y.~Mo and B.~Sinopoli, ``Towards finding the critical value for {K}alman
  filtering with intermittent observations,'' \emph{arXiv preprint
  arXiv:1005.2442}, 2010.

\bibitem{sinopoli2005optimal}
B.~Sinopoli, L.~Schenato, M.~Franceschetti, K.~Poolla, and S.~S. Sastry,
  ``Optimal control with unreliable communication: the {TCP} case,'' in
  \emph{Proceedings of the American Control Conference}, 2005, pp. 3354--3359.

\bibitem{imer2006optimalautomatica}
O.~C. Imer, S.~Y{\"u}ksel, and T.~Ba{\c{s}}ar, ``Optimal control of {LTI}
  systems over unreliable communication links,'' \emph{Automatica}, vol.~42,
  no.~9, pp. 1429--1439, 2006.

\bibitem{sinopoli2008optimal}
B.~Sinopoli, L.~Schenato, M.~Franceschetti, K.~Poolla, and S.~Sastry, ``Optimal
  linear {LQG} control over lossy networks without packet acknowledgment,''
  \emph{Asian Journal of Control}, vol.~10, no.~1, pp. 3--13, 2008.

\bibitem{foundation-ncs-packet-drops}
L.~Schenato, B.~Sinopoli, M.~Franceschetti, K.~Poolla, and S.~Sastry,
  ``Foundations of control and estimation over lossy networks,''
  \emph{Proceedings of the IEEE}, vol.~95, pp. 163--187, 2007.

\bibitem{gilbert1960capacity}
E.~N. Gilbert, ``Capacity of a burst-noise channel,'' \emph{Bell system
  technical journal}, vol.~39, no.~5, pp. 1253--1265, 1960.

\bibitem{elliott1963estimates}
E.~Elliott, ``Estimates of error rates for codes on burst-noise channels,''
  \emph{Bell system technical journal}, vol.~42, no.~5, pp. 1977--1997, 1963.

\bibitem{huang2006cdc}
M.~Huang and S.~Dey, ``Kalman filtering with {M}arkovian packet losses and
  stability criteria,'' in \emph{Proceedings of the 45th IEEE Conference on
  Decision and Control}, 2006, pp. 5621--5626.

\bibitem{huang-dey-stability-kf}
------, ``Stability of {K}alman filtering with {M}arkovian packet losses,''
  \emph{Automatica}, vol.~43, pp. 598--607, 2007.

\bibitem{xie2007peak}
L.~Xie and L.~Xie, ``Peak covariance stability of a random {R}iccati equation
  arising from {K}alman filtering with observation losses,'' \emph{Journal of
  Systems Science and Complexity}, vol.~20, no.~2, pp. 262--272, 2007.

\bibitem{xie2008stability}
------, ``Stability of a random {R}iccati equation with {M}arkovian binary
  switching,'' \emph{IEEE Transactions on Automatic Control}, vol.~53, no.~7,
  pp. 1759--1764, 2008.

\bibitem{you2011mean}
K.~You, M.~Fu, and L.~Xie, ``Mean square stability for {K}alman filtering with
  {M}arkovian packet losses,'' \emph{Automatica}, vol.~47, no.~12, pp.
  2647--2657, 2011.

\bibitem{shi-tac10}
L.~Shi, M.~Epstein, and R.~M. Murray, ``Kalman filtering over a packet-dropping
  network: A probabilistic perspective,'' \emph{IEEE Transactions on Automatic
  Control}, vol.~55, no.~9, pp. 594--604, 2010.

\bibitem{yilin12criticalvalue}
Y.~Mo and B.~Sinopoli, ``Kalman filtering with intermittent observations: Tail
  distribution and critical value,'' \emph{IEEE Transactions on Automatic
  Control}, vol.~57, no.~3, pp. 677--689, March 2012.

\bibitem{kar2012kalman}
S.~Kar, B.~Sinopoli, and J.~M. Moura, ``Kalman filtering with intermittent
  observations: Weak convergence to a stationary distribution,'' \emph{IEEE
  Transactions on Automatic Control}, vol.~57, no.~2, pp. 405--420, 2012.

\bibitem{censi2011kalman}
A.~Censi, ``Kalman filtering with intermittent observations: convergence for
  semi-{M}arkov chains and an intrinsic performance measure,'' \emph{IEEE
  Transactions on Automatic Control}, vol.~56, no.~2, pp. 376--381, 2011.

\bibitem{xie2012stochastic}
L.~Xie, ``Stochastic comparison, boundedness, weak convergence, and ergodicity
  of a random {R}iccati equation with {M}arkovian binary switching,''
  \emph{SIAM Journal on Control and Optimization}, vol.~50, no.~1, pp.
  532--558, 2012.

\bibitem{durrett2010probability}
R.~Durrett, \emph{Probability: Theory and Examples}.\hskip 1em plus 0.5em minus
  0.4em\relax Cambridge university press, 2010.

\bibitem{horn2012matrix}
R.~A. Horn and C.~R. Johnson, \emph{Matrix analysis}.\hskip 1em plus 0.5em
  minus 0.4em\relax Cambridge university press, 2012.

\bibitem{xiao2009kalman}
N.~Xiao, L.~Xie, and M.~Fu, ``Kalman filtering over unreliable communication
  networks with bounded {M}arkovian packet dropouts,'' \emph{International
  Journal of Robust and Nonlinear Control}, vol.~19, no.~16.

\bibitem{wang1995finite}
H.~S. Wang and N.~Moayeri, ``Finite-state {M}arkov channel-a useful model for
  radio communication channels,'' \emph{IEEE Transactions on Vehicular
  Technology}, vol.~44, no.~1, pp. 163--171, 1995.

\bibitem{zhang1999finite}
Q.~Zhang and S.~A. Kassam, ``Finite-state markov model for rayleigh fading
  channels,'' \emph{IEEE Transactions on Communications}, vol.~47, no.~11, pp.
  1688--1692, 1999.

\bibitem{sinopoli}
B.~Sinopoli, L.~Schenato, M.~Franceschetti, K.~Poolla, M.~Jordan, and
  S.~Sastry, ``Kalman filtering with intermittent observations,'' \emph{IEEE
  Transactions on Automatic Control}, vol.~49, no.~9, pp. 1453--1464, 2004.

\bibitem{solo91}
V.~Solo, ``One step ahead adaptive controller with slowly time-varying
  parameters,'' Department of EECS, John Hopkins University, Baltimore, MD,
  USA, Tech. Rep., 1991.

\bibitem{Costa04TAC}
O.~L.~V. Costa and M.~Fragoso, ``Comments on "stochastic stability of jump
  linear systems",'' \emph{IEEE Transactions on Automatic Control}, vol.~49,
  no.~8, pp. 1414--1416, Aug 2004.

\bibitem{lancaster1995ARE}
P.~Lancaster and L.~Rodman, \emph{Algebraic {R}iccati equations}.\hskip 1em
  plus 0.5em minus 0.4em\relax Oxford University Press, 1995.

\end{thebibliography}

\medskip

\medskip

\medskip

\noindent {\sc Junfeng Wu and Karl H. Johansson} \\
{\noindent {\small  ACCESS Linnaeus Centre,
   School of Electrical Engineering,
KTH Royal Institute of Technology,
\\ Stockholm 100 44, Sweden }\\}
       {\small Email: {\tt\small junfengw@kth.se, kallej@kth.se}

\medskip

{\noindent {\sc Guodong Shi and Brian D. O.  Anderson}} \\
{\noindent {\small College of Engineering and Computer Science, The Australian National University, \\ Canberra, ACT 0200 Australia}}\\  {\small Email: } {\tt\small guodong.shi@anu.edu.au, brian.anderson@anu.edu.au}

\end{document}